\documentclass[letterpaper,twocolumn,aps,prl,floatfix,superscriptaddress,footinbib,preprintnumbers]{revtex4-2}

\usepackage{float}	
\usepackage[caption=false]{subfig}
\usepackage{graphicx}	
\usepackage{mathtools}
\usepackage{amsfonts}
\usepackage{physics}
\usepackage{slashed}
\usepackage{circuitikz}
\usepackage{import}
\usepackage[colorlinks=true]{hyperref}
\usepackage[capitalise]{cleveref}

\graphicspath{ {./figs/}}

\begin{document}	
	\title{
		High-Energy Collision of Quarks and Mesons in the Schwinger Model:
		\\
		From Tensor Networks to Circuit QED}
	\date{\today}
	\author{Ron Belyansky}
	\email{rbelyans@umd.edu}
	\author{Seth Whitsitt}
	\affiliation{Joint Center for Quantum Information and Computer Science, NIST/University of Maryland, College Park, MD 20742 USA}
	\affiliation{Joint Quantum Institute, NIST/University of Maryland, College Park, MD 20742 USA}
	\author{Niklas~Mueller}
	\affiliation{InQubator for Quantum Simulation (IQuS), Department of Physics, University of Washington, Seattle, WA 98195, USA}
	\author{Ali Fahimniya}
	\author{Elizabeth R. Bennewitz}
	\affiliation{Joint Center for Quantum Information and Computer Science, NIST/University of Maryland, College Park, MD 20742 USA}
	\affiliation{Joint Quantum Institute, NIST/University of Maryland, College Park, MD 20742 USA}
	\author{Zohreh~Davoudi}
	\affiliation{Maryland Center for Fundamental Physics and Department of Physics, University of Maryland, College Park, MD 20742 USA}
	\affiliation{Joint Center for Quantum Information and Computer Science, NIST/University of Maryland, College Park, MD 20742 USA}
	\author{Alexey V. Gorshkov}
	\affiliation{Joint Center for Quantum Information and Computer Science, NIST/University of Maryland, College Park, MD 20742 USA}
	\affiliation{Joint Quantum Institute, NIST/University of Maryland, College Park, MD 20742 USA}
	
	\begin{abstract}
		With the aim of studying nonperturbative out-of-equilibrium dynamics of high-energy particle collisions on quantum simulators, we investigate the scattering dynamics of lattice quantum electrodynamics in 1+1 dimensions. 
		Working in the bosonized formulation of the model
		and in the thermodynamic limit, we use uniform-matrix-product-state tensor networks to construct multi-particle wave-packet states, evolve them in time, and detect outgoing particles post collision.
		This facilitates the numerical simulation of scattering experiments in both confined and deconfined regimes of the model at different energies, giving rise to rich phenomenology, including inelastic production of quark and meson states, meson disintegration,
		and dynamical string formation and breaking.
		We obtain elastic and inelastic scattering cross sections, together with time-resolved momentum and position distributions of the outgoing particles. 
		Furthermore, we propose an analog circuit-QED implementation of the scattering process that is native to the platform, requires minimal ingredients and approximations, and enables practical schemes for particle wave-packet preparation and evolution.
		This study highlights the role of classical and quantum simulation in enhancing our understanding of scattering processes in quantum field theories in real time.
	\end{abstract}
	\maketitle
	
	{\it Introduction.---}Scattering processes in nuclear and high-energy physics play an essential role in studies of fundamental particles and interactions.
	Current and future frontiers in scattering experiments are the Large Hadron Collider, the Relativistic Heavy-Ion Collider~\cite{florkowski2010phenomenology,lovato2022long}, the Electron-Ion Collider~\cite{accardiElectronIonColliderNext2016,Achenbach:2023pba}, and the Deep Underground Neutrino Experiment~\cite{gallagher2011neutrino,alvarez2018nustec,kronfeld2019lattice,Ruso:2022qes}.
	Collisions in these experiments involve hadronic initial states and complex many-particle final states. The scattering proceeds in a multi-stage process and may encompass a wide range of phenomena, including the formation of exotic matter~\cite{sorensen2023dense,lovato2022long}, such as quark-gluon plasma~\cite{bass1999signatures,Shuryak:1980tp}, thermalization~\cite{baier2001bottom,berges2021qcd}, fragmentation~\cite{Andersson:1983ia,Webber:1983if}, and  hadronization~\cite{bass2000dynamics,andronic2018decoding}. 
	Ideally, such phenomenology should be grounded in first-principles quantum-chromodynamics (QCD) descriptions. While perturbation theory and QCD factorization~\cite{bjorken1969asymptotic,gross1973ultraviolet,collins1989factorization,blumlein2013theory}, as well as lattice QCD~\cite{kronfeld2022lattice,davoudi2022report,beane2011nuclear,davoudi2021nuclear,drischler2021towards,ding2015thermodynamics,ratti2018lattice,usqcd2019hot,guenther2021overview}, have brought about impressive advances, a full understanding of scattering processes in QCD is still lacking. In particular, lattice-QCD studies of hadronic scattering based on Monte-Carlo methods in Euclidean spacetime have so far been limited to low energy and low inelasticities~\cite{Briceno:2017max,Hansen:2019nir}. These also do not track the state evolution after the collision.
	
	First-principles simulations of high-energy particle scattering are a prime application for quantum simulators~\cite{jordan2011quantum,jordanQuantumAlgorithmsQuantum2012,jordan2018bqp,roggero2020quantum,mueller2020deeply,barata2021single,farrell2023preparations,surace2021scattering,Bauer:2022hpo,Beck:2023xhh,klco2022standard,bauerQuantumSimulationFundamental2023}.
	However, realistic experiments involve a vast range of spatial and temporal scales, placing their simulation beyond the capabilities of current digital quantum computers.
	Analog quantum simulators may enable simulating larger Hilbert spaces and longer times, 
	but concrete proposals for simulating scattering processes in quantum field theories are lacking.
	At the same time, classical tensor-network methods have been shown to successfully capture ground-state~\cite{Schollwock2011}, and to some degree dynamical~\cite{paeckelTimeevolutionMethodsMatrixproduct2019}, phenomena in gapped theories, including scattering processes~\cite{pichlerRealtimeDynamicsLattice2016,rigobelloEntanglementGenerationQED2021a,vandammeRealtimeScatteringInteracting2021,milstedCollisionsFalseVacuumBubble2022}, particularly in $1+1$ dimensions, but their reach remains limited in simulating general scattering problems in quantum field theories. This manuscript advances both analog quantum simulation and tensor-network-based classical simulation for a prototypical model of QCD, the lattice Schwinger model, i.e., lattice quantum electrodynamics (QED) in 1+1 dimensions.
	Previous tensor-network~\cite{buyensMatrixProductStates2014, buyensConfinementStringBreaking2016,buyensRealtimeSimulationSchwinger2017,byrnesDensityMatrixRenormalization2002, banuls2013mass, rico2014tensor, pichlerRealtimeDynamicsLattice2016, banulsThermalEvolutionSchwinger2015,zappTensorNetworkSimulation2017,rigobelloEntanglementGenerationQED2021a, funckeTopologicalVacuumStructure2020, buttTensorNetworkFormulation2020, banuls2020review,meurice2022tensor} and quantum-simulation~\cite{martinez2016real,klco2018quantum,nguyenDigitalQuantumSimulation2022,muellerQuantumComputationDynamical2022,chakraborty:2020uhf,de2022quantum,shawQuantumAlgorithmsSimulating2020,kan2021lattice,zhouThermalizationDynamicsGauge2022,yangObservationGaugeInvariance2020,milScalableRealizationLocal2020,banerjeeAtomicQuantumSimulation2012,haukeQuantumSimulationLattice2013a,wieseUltracoldQuantumGases2013,zoharQuantumSimulationsLattice2015,yangAnalogQuantumSimulation2016,davoudiAnalogQuantumSimulations2020,luoFrameworkSimulatingGauge2020a,notarnicolaRealtimedynamicsQuantumSimulation2020,suraceLatticeGaugeTheories2020,davoudiSimulatingQuantumField2021,andrade2022engineering,marcosSuperconductingCircuitsQuantum2013a,halimehTuningTopologicalAngle2022,osborneSpinMathrmUQuantum2023a,zhangObservationMicroscopicConfinement2023} studies of the model focused on formulations involving fermion (or qubit) degrees of freedom (with or without gauge fields).
	Motivated to address, more generally, theories with bosonic content, here we instead consider the bosonic dual of the theory, a particular type of a massive Sine-Gordon model.

	Our first objective is a numerical exploration of high-energy real-time scattering phenomenology in the model.
	We work in the nonperturbative regime, near the confinement-deconfinement critical point and in the thermodynamic limit, using uniform matrix product states (uMPS)~\cite{vanderstraetenTangentspaceMethodsUniform2019}, which allows for the construction~\cite{vandammeRealtimeScatteringInteracting2021,milstedCollisionsFalseVacuumBubble2022} and collision of numerically-exact quasiparticle wave packets in the interacting theory at various energies, resulting in nontrivial inelastic effects. In contrast, earlier works were limited to elastic scattering at either weak (nearly free fermions)~\cite{rigobelloEntanglementGenerationQED2021a} or strong (nearly free bosons)~\cite{pichlerRealtimeDynamicsLattice2016} coupling regimes. We focus on  spatial, temporal, and momentum-resolved diagnostics of elastic and inelastic processes of quark and meson states, involving phenomena such as meson disintegration, dynamical string formation and breaking, and the creation of quark and (excited) meson states. We also investigate the role of entanglement in high-energy scattering~\cite{pichlerRealtimeDynamicsLattice2016,Berges:2017zws,Berges:2017hne,kharzeev2017deep,hagiwara2018classical,kovner2019entanglement,beane2019entanglement,beane2020chiral,beane2021geometry}.
	Our second objective is to propose an analog circuit-QED implementation of the bosonized lattice Schwinger model. 
	Recently, the bosonic dual was shown to be approximately realizable by circular Rydberg states~\cite{kruckenhauserHighdimensionalSymmetricRydberg2022}.
	In contrast, we will show that circuit QED's basic components, its native bosonic degrees of freedom, and the available ultrastrong coupling~\cite{Forn-Diaz2019, FriskKockum} allow the model to be implemented in a simple circuit with minimal ingredients and approximations, making it particularly suitable for near-term quantum simulation that goes beyond the classical simulation methods.

	{\it Model.---}The massive Schwinger model has the Lagrangian density 
	\begin{equation}
		\label{eq:Schwinger-lagr}
		\mathcal{L} = \bar{\psi} \big(i\gamma^\mu {\partial}_\mu-e\gamma^\mu{A}_{\mu}-m \big) \psi -\frac{1}{4}F_{\mu\nu}F^{\mu\nu},
	\end{equation}
	where $\psi(x,t)$ is a 2-component Dirac spinor, $\gamma^0 = \sigma^z, \gamma^1 = i \sigma^y$ with $\sigma^z,\sigma^y$ being the Pauli matrices, $m$ is the mass, $e$ is the electric charge, and $A_{\mu}(x,t)$ and $F_{\mu\nu}(x,t)$ are the gauge field and the field-strength tensor, respectively. 
	\Cref{eq:Schwinger-lagr} is dual to a bosonic scalar field theory with the Hamiltonian~\cite{colemanChargeShieldingQuark1975,colemanMoreMassiveSchwinger1976}
	\begin{equation}
		\label{eq:Schwinger-bose-ham}
		H =  \int dx \, :\bigg[\frac{\Pi^2}{2} +\frac{(\partial_x\phi)^2}{2}+\frac{e^2\phi^2}{2\pi}-\frac{b\,m\,e\cos(\sqrt{4\pi}\phi-\theta)}{2\pi^{3/2}} \bigg]:,
	\end{equation}
	where normal-ordering ($::$) is with respect to $e/\sqrt{\pi}$, $\phi(x)$ and $\Pi(x)$ are the scalar field and conjugate momentum, respectively, $b=\exp(\gamma)$ with $\gamma$ being Euler's constant, and $\theta \in (-\pi,\pi]$, with its origin explained in Ref.~\cite{colemanMoreMassiveSchwinger1976} and the Supplemental Material (SM)~\cite{supp} (we assume $\hbar=c=1$ throughout, where $c$ is the speed of light). 
	We work with a lattice regularization of \cref{eq:Schwinger-bose-ham} given by 
	\begin{equation}
		\label{eq:Schwinger-bose-lattice}
		H = \chi\sum_x \bigg[\frac{\pi_x^2}{2}+\frac{(\phi_{x}-\phi_{x-1})^2}{2} +\frac{\mu^2\phi_x^2}{2}-\lambda\cos(\beta\phi_x-\theta)\bigg],
	\end{equation}
	where $x$ labels lattice sites, $\comm{\phi_x}{\pi_y}=i\delta_{xy}$, $\chi=1/a$, $\beta=\sqrt{4\pi}$, $\mu^2=a^2e^2/\pi$, $\lambda = a^2\,b\, m\,e \exp[2\pi\Delta(a)]/2\pi^{3/2}$, $a$ is the lattice spacing, and $\Delta(a)$ is the lattice Feynman propagator at the origin \cite{colemanQuantumSineGordonEquation1975,ohataMonteCarloStudy2023}. We set $a=1$, with the continuum limit corresponding to $\mu,\lambda \to 0$. 
	Quantities are assumed in lattice units throughout.

	\begin{figure}[t!]
		\centering
		\includegraphics[width=\linewidth]{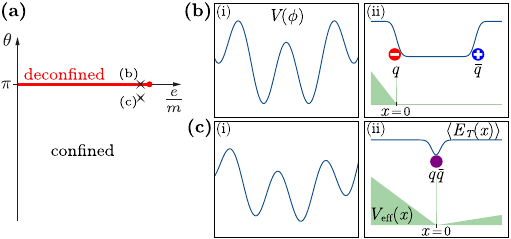}
		\caption{(a) Sketch of the phase diagram of the massive Schwinger model as a function of $e/m$ (corresponding to $\mu/\lambda$) and $\theta$. The red dot is the Ising critical point, where the deconfined phase (red line) terminates. Points (b) and (c) correspond to the two regimes considered in the main text. Panels (b,i) and (c,i) show the corresponding scalar potential $V(\phi)=\frac{1}{2}\mu^2\phi^2-\lambda \cos(\sqrt{4\pi}\phi-\theta)$ [\cref{eq:Schwinger-bose-lattice}]. Panels (b,ii) and (c,ii) show both the effective potential between the quarks [\cref{eq:H_eff-quark}] (green) and the electric/scalar-field distributions (blue) due to the quarks and mesons.}
		\label{fig:phase-diagram}
	\end{figure}

	To gain insight into the anticipated phenomenology, we proceed with a numerical study of the collision dynamics in the lattice Schwinger model. 
	While quantitative predictions for the continuum theory require an extrapolation procedure \cite{zappTensorNetworkSimulation2017,zacheContinuumLimitMathrmD2022}, here only fixed, but sufficiently small, values of $\mu$ and $\lambda$ are considered.
	The model has two dimensionless parameters, the ratio $e/m$, corresponding to $\mu/\lambda$ in \cref{eq:Schwinger-bose-lattice}, and the angle $\theta$ representing a constant background electric field $E_\theta = \frac{e}{2\pi}\theta$. 
	Gauss's law, $\partial_xE=e\psi^\dagger\psi$, ties the total electric field $E_T=E_\theta+E$ to the dynamical charges, and equals  $E_T=\frac{e}{\sqrt{\pi}}\phi$ in the bosonic dual~\cite{shankarDeconfinementAsymptoticHalfasymptotic2005c}.

	We study two regimes near the $\mathbb{Z}_2$ critical point, shown in \cref{fig:phase-diagram} as (b) and (c). Point (b) is in the deconfined phase [red line at $\theta=\pi$ in \cref{fig:phase-diagram}(a) terminating at the Ising critical point], where 
	the ground state is two-fold degenerate [\cref{fig:phase-diagram}(b,i)]. Here, fundamental excitations are  ``half-asymptotic" \cite{colemanMoreMassiveSchwinger1976} fermions (``quarks"), appearing as topological kinks in the bosonic dual [see \cref{fig:phase-diagram}(b,ii)].
	Point (c) in \cref{fig:phase-diagram}(a) is in the confined phase, with a unique ground state [\cref{fig:phase-diagram}(c,i)] and quark-antiquark bound-state (``meson") excitations.

	{\it Quark-antiquark scattering.---}
	Constructing a uMPS representation of the two ground states~\cite{zauner-stauberVariationalOptimizationAlgorithms2018} in the deconfined phase [\cref{fig:phase-diagram}(b)], we use the uMPS quasiparticle ansatz~\cite{haegemanVariationalMatrixProduct2012,haegemanElementaryExcitationsGapped2013} to obtain single-particle energy-momentum eigenstates with dispersion $\mathcal{E}(p)$ and momenta $p\in [-\pi,\pi)$ \cite{supp}.  
	From this, we construct two Gaussian wave packets, localized in momentum and position space, centered at opposite momenta $\pm p_0$. 
	The initial state consists of a finite nonuniform region of 150 to 300 sites containing the two wave packets, and
	is surrounded by the uniform vacuum [we choose the vacuum with positive $E_T$, i.e., the right minimum of~\cref{fig:phase-diagram}(b,i)]. 
	We then time-evolve this state under the Hamiltonian in \cref{eq:Schwinger-bose-lattice}, while dynamically expanding the nonuniform region~\cite{milstedVariationalMatrixProduct2013b,phienDynamicalWindowsRealtime2013,zaunerTimeEvolutionComoving2015} up to 
	600 to 1300 sites (see SM \cite{supp}).
	By working near the critical point, where the quark mass $m_q\equiv \mathcal{E}(p=0)$ is small, one can consider momenta up to $\abs{p_0}\lesssim 0.8$. These are sufficiently small to keep the physics in the long-wavelength regime of the lattice model, where the dispersion is approximately relativistic $\mathcal{E}(p)\approx({p^2+m_q^2})^{\frac{1}{2}}$, but highly relativistic center-of-mass (c.m.) energies $\mathcal{E}_{\text{c.m.}}\equiv 2\mathcal{E}(p_0)\lesssim 30m_q$ are achieved.

	\begin{figure}[t!]
		\centering
		\includegraphics[width=\linewidth]{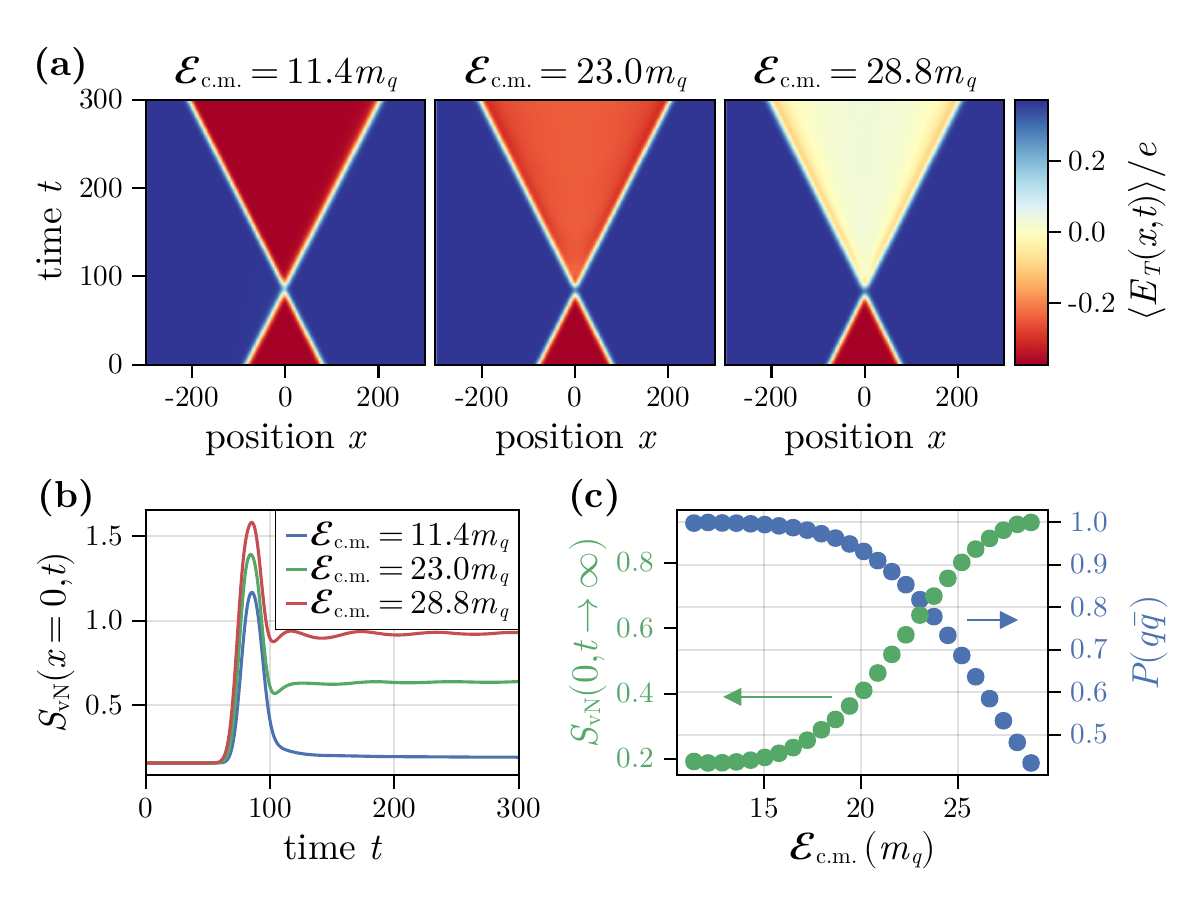}
		\caption{Quark-antiquark scattering in the deconfined phase. (a) Time evolution of the electric field for different center-of-mass energies.  
			(b) Time evolution of the von Neumann entanglement entropy for a cut at $x = 0$, for the same three collisions as in (a). (c) Elastic scattering probability (right, blue) and asymptotic 
			von Neumann entanglement entropy for the $x=0$ cut (left, green) as a function of the center-of-mass energy. The parameters are $\mu^2=0.1$ and $\lambda=0.5$ [see \cref{eq:Schwinger-bose-lattice}].} 
		\label{fig:kink_scat}
	\end{figure} 
	
	\Cref{fig:kink_scat}(a) shows the space-time distribution of the electric field for collisions at three representative energies, $\mathcal{E}_{\text{c.m.}}/m_q= $ 11.4, 23.0, and 28.8. 
	Initially, the quark and antiquark are  separated, resembling \cref{fig:phase-diagram}(b,ii), with electric field between the charges equal in magnitude but opposite in sign to the field outside [the two regions correspond to the two degenerate ground states in \cref{fig:phase-diagram}(b,i)].
	Under time evolution, the two charges propagate ballistically, shrinking the negative-field region until they collide. 
	During the collision, the particles bounce off each other and reverse their propagation direction elastically, the sole process at lower energies. 
	Specifically, as can be seen in \cref{fig:kink_scat}(a), at the lowest energy, $\mathcal{E}_{\text{c.m.}}/m_q= $ 11.4,  the post-collision value of $E_T$ between the charges is practically equal to the pre-collision value.
	For the higher-energy collisions, $\mathcal{E}_{\text{c.m.}}/m_q= $ 23.0 and 28.8, 
	an increase of the post-collision electric field is observed, signalling additional charge production.

	While our numerical approach does not rely on strong- or weak-coupling expansion, the relevant scattering channels can be understood from weak-coupling arguments as follows.
	In the SM \cite{supp}, we derive, in the nonrelativistic limit, an effective potential between opposite charges at the lowest order in $e/m$ starting from \cref{eq:Schwinger-lagr}, which reads (in the center-of-mass frame) 
	\begin{equation}
		\label{eq:H_eff-quark}
		V_{\text{eff}}(x)= \frac{e^2}{2}\qty(\abs{x}-\frac{\theta}{\pi}x)+\frac{e^2}{4m^2}\delta(x)\,.
	\end{equation}
	Here, $x$ is the distance between charges.
	For $\theta\neq\pi$, one recovers linear confinement [\cref{fig:phase-diagram}(c,ii)]~\cite{colemanMoreMassiveSchwinger1976,rotheScreeningConfinement1979,shankarDeconfinementAsymptoticHalfasymptotic2005c,buyensConfinementStringBreaking2016}, while at $\theta=\pi$, charges experience short-range \emph{repulsion} due to the delta function in \cref{eq:H_eff-quark} [\cref{fig:phase-diagram}(b,ii)]. 
	This implies the absence of stable bound states (mesons) in the deconfined phase, which is confirmed numerically in the SM~\cite{supp}.
	All possible scattering channels are, therefore, (even-numbered) multi-quark states.
	The lowest-order inelastic channel is the four-quark production ($q\bar{q}\rightarrow q\bar{q}q\bar{q}$), exhibiting quark fragmentation.
	The two inner particles screen the electric field produced by the outer two, consistent with the two rightmost panels in \cref{fig:kink_scat}(a).
	\begin{figure*}[ht]
		\centering
		\includegraphics[width=\linewidth]{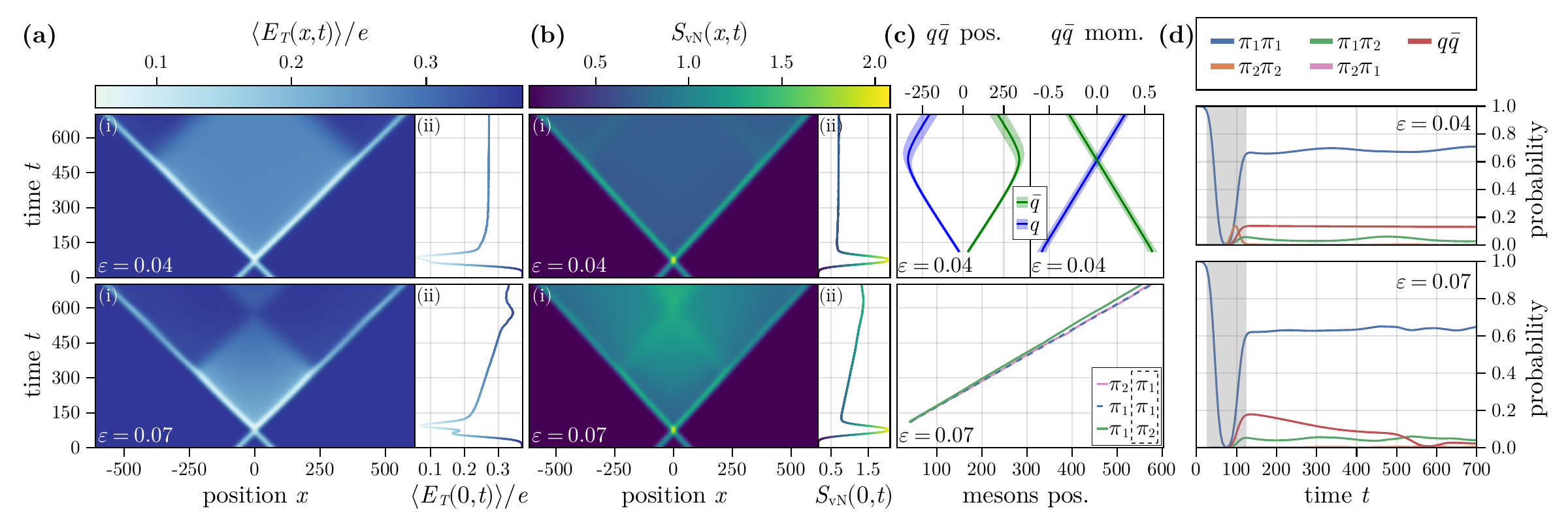}
		\caption{Meson-meson scattering in the confined phase. (a) Time evolution of the electric field for different $\theta=\pi-\varepsilon$ at all positions $x$ [panels (i)] and at $x=0$ [panels (ii)] with $\mu^2=0.1$ and $\lambda=0.5$ as in \cref{fig:kink_scat}. The wave packets are centered at $p_0=\pm0.6$, corresponding to $\mathcal{E}_{\text{c.m.}}/{m_{\pi_{1}}}=6.84, 5.95$ for $\varepsilon=0.04, 0.07$. 
			(b) 
			Time evolution of the von Neumann entanglement entropy  at all positions $x$ [panels (i)] and at $x=0$ [panels (ii)]. 
			(c) Momenta and positions (mean $\pm$ std.~extracted from a Gaussian fit of the projected distributions) of the quarks for $\varepsilon=0.04$ (top) and the mean positions of the right-moving mesons for $\varepsilon=0.07$ (bottom).  
			(d) Probabilities of two-particle states $\mu\nu$ ($\mu,\nu \in [\pi_{1},\pi_{2},q,\bar{q}]$) where $\mu/\nu$ is the particle on the left/right. The curves for $\pi_{1}\pi_{2}$ and $\pi_{2}\pi_{1}$ overlap due to the reflection symmetry of the initial state. Near the initial collision (shaded region), and the secondary collision at $t\approx550$ for $\varepsilon=0.07$, the state cannot be fully captured by a basis of asymptotic particles.}
		
		\label{fig:meson_scat}
	\end{figure*} 

	Elastic and inelastic processes are also distinguished by the von Neumann entanglement entropy \footnote{$S_\mathrm{vN}(x,t)=-\tr(\rho_{>x}(t) \ln \rho_{>x}(t))$ with $\rho_{>x}(t)$ being the reduced density matrix for sites $y > x$} across the collision point ($x=0$), shown in \cref{fig:kink_scat}(b) as a function of time.  \Cref{fig:kink_scat}(c) also shows the asymptotic ($t \to \infty$) entanglement as a function of the collision energy. The entanglement is maximal during the collision but quickly approaches a constant afterwards. At lower energies, it nearly returns to its pre-collision (vacuum) value. 
	A small increase is observed because different momentum components of the wave packets acquire slightly different elastic scattering phase shifts~\cite{milstedCollisionsFalseVacuumBubble2022}.
	At higher energies, however, significant net entanglement is generated, indicating inelastic particle production~\cite{jordanQuantumAlgorithmsQuantum2012}.

	Finally, we compute elements of the momentum-resolved scattering S-matrix by projecting the post-collision state onto a basis of asymptotic two-particle states (see SM \cite{supp}). This basis is constructed from the single-particle wavefunctions, requiring the particles to be widely separated to ensure orthogonality and avoid interaction effects.
	For $2\rightarrow 2$ scattering, this is guaranteed sufficiently far from the collision point. 
	From this, we obtain the elastic scattering probability $P(q\bar{q})$, displayed in \cref{fig:kink_scat}(c), as a function of the collision energy.

	The elastic scattering probability is near unity at lower energies, decreasing monotonically, falling below $0.5$ around $\mathcal{E}_{\text{c.m.}}/m_q\gtrsim 28$.
	Interestingly, the energy required for significant inelastic scattering is many times the threshold energy ($\mathcal{E}_{\text{\text{c.m.}}}=4m_q$).
	While we did not obtain the precise contribution of the four-quark (or higher-quark-number) states \footnote{Projecting the state on four widely-separated quarks basis states resulted in a negligible contribution.
		This suggests that the four particles are not sufficiently spatially separated.},
	the decrease of $P(q\bar{q})$ confirms the presence of significant inelastic scattering,  consistent with the increase in entanglement in \cref{fig:kink_scat}(b) and the screening of $E_T$ in \cref{fig:kink_scat}(a).

	{\it Meson-meson scattering.---}
	We next choose $\theta = \pi-\varepsilon$ with $\varepsilon\ll 1$, which gives rise to weak confinement of quarks, but keeps us close to the critical point (all other parameters are unchanged). 
	In contrast to the deconfined regime, the interplay of high-energy and weak confinement yields rich behavior following the collision. 
	There are multiple stable scalar meson excitations, which are labeled by $\pi_{j}\, (j=1,2,...)$, with increasing masses $m_{\pi_{j}}$. Here, we consider $\pi_{1}\pi_{1}$ collisions, with meson wave packets prepared similarly as before, centered at $p_0=\pm0.6$ with $\mathcal{E}_{\text{c.m.}}/{m_{\pi_{1}}}=6.84$ (5.95) for $\varepsilon=0.04$ (0.07).
	
	The electric-field evolution for the two collisions is displayed in \cref{fig:meson_scat}(a,i).  
	Before the collision, the background electric field is only locally disturbed by the charge-neutral mesons [\cref{fig:phase-diagram}(c,ii)]. 
	After the collision, the mesons partially fragment into a quark-antiquark pair. The quarks are joined by an electric-field string 
	which screens the background electric field (light-blue regions) inside the collision cone.
	As the quarks travel outward, their kinetic energy gets converted into the potential energy of the string. Eventually, they turn and propagate back in the opposite direction [see also \cref{fig:meson_scat}(c)] causing a second collision. Weaker confinement $\varepsilon=0.04$ allows the quarks to propagate farther.

	Next, we project the time-evolved state onto two-particle components, focusing on the lightest two mesons $\pi_1,\pi_2$, and the quark-antiquark pair $q\bar{q}$. While the latter are not true (i.e., asymptotic) quasiparticles, at weak confinement $\varepsilon\ll 1$, (anti)quarks can be approximately described by the modified quasiparticle ansatz of Ref.~\cite{milstedCollisionsFalseVacuumBubble2022}.
	This requires a uMPS representation of the electric-flux string, which we approximate by its lowest energy state, a so-called ``false-vacuum" state \cite{colemanFateFalseVacuum1977,callanFateFalseVacuum1977}, corresponding to the second (local) minimum in \cref{fig:phase-diagram}(c,i).

	\Cref{fig:meson_scat}(d) shows the probabilities of the $\pi_{1}\pi_{1}$, $\pi_{2}\pi_{2}$, $\pi_{1}\pi_{2}$, $\pi_{2}\pi_{1}$, and $q\bar{q}$ states (where in state $\mu \nu$, the particle $\mu/\nu$ is on the left/right). 
	One can observe significant flavor-conserving elastic scattering, $\pi_{1}\pi_{1}\rightarrow \pi_{1}\pi_{1}$, a smaller probability of exciting one of the outgoing mesons,  $\pi_{2}\pi_{1}$ and $\pi_{1}\pi_{2}$, and a substantial $q\bar{q}$ component.
	Interestingly, for $\varepsilon=0.07$, the $q\bar{q}$ component is decreasing in time, indicating string breaking~\cite{buyensConfinementStringBreaking2016, hebenstreitRealTimeDynamicsString2013}, which is also visible in the gradual increase of the bipartite entanglement entropy in \cref{fig:meson_scat}(b,i) [see also \cref{fig:meson_scat}(b,ii)], and in the gradual reduction of the electric-field screening [\cref{fig:meson_scat}(a,ii)].
	At a late time $t=700$, asymptotic two-particle states (including the quark-antiquark state) account for about $90\,\%$ ($76\,\%$) of the state at $\varepsilon=0.04$ (0.07)~\footnote{We verified that the missing wavefunction weight is not accounted for by three or four widely-separated particle basis states.}.

	The projection onto the asymptotic two-particle basis also provides the full momentum, and consequently position, distributions of the particles. \Cref{fig:meson_scat}(c) shows the mean and standard deviation of the positions and momenta of the quarks, and the mean positions of the mesons, computed from fits of these distributions to a Gaussian form. 
	The mean momenta of the quarks are approximately $\expval{p(t)}\propto \pm t$, in agreement with the  expectation from the linear potential of \cref{eq:H_eff-quark}. Their extracted positions in \cref{fig:meson_scat}(c) are consistent with the boundaries of the screened-field region in \cref{fig:meson_scat}(a,i) and with the localized increase in the entanglement entropy in \cref{fig:meson_scat}(b,i).
	From the mean position of the mesons, \cref{fig:meson_scat}(c), one can see that the heavier meson $\pi_2$ has a slightly lower average velocity compared to $\pi_1$, as expected.

\begin{figure}[tb!]	
		\centering
		\includegraphics[width=\linewidth]{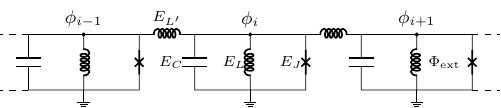}
		\caption{Lumped-element circuit diagram that realizes \cref{eq:Schwinger-bose-lattice}.}
		\label{fig:circuit-diag}
	\end{figure}

	{\it Circuit-QED implementation.---}
	The increasingly large entanglement production occurring with higher-energy collisions [\cref{fig:kink_scat}] or stronger confinement  and longer collision times [\cref{fig:meson_scat}] limits the regimes of applicability of the MPS-based methods. This motivates the use of quantum simulators as an alternative approach, which is expected to evade such limitations.
	Remarkably, the lattice Schwinger Hamiltonian [\cref{eq:Schwinger-bose-lattice}] can be exactly realized in a simple superconducting circuit, shown in \cref{fig:circuit-diag}. The circuit can be regarded as a chain of inductively coupled fluxoniums \cite{ozgulerExcitationDynamicsInductively2021}.
	It consists of nodes $i$, each corresponding to a lattice site with a local bosonic degree of freedom described by flux $\phi_i$ and charge $\pi_i$, composed of a parallel arrangement of a capacitor, an inductor, and a Josephson junction with respective energies $E_C,E_L$, and $E_J$~\cite{blaisCircuitQuantumElectrodynamics2021}. Nodes are coupled by inductors with energy $E_{L'}$. The circuit parameters are related to those of \cref{eq:Schwinger-bose-lattice} via $\chi = \frac{8E_C}{\beta^2}$, $\frac{E_{L'}\beta^4}{8E_C}=1$, $\mu^2=\frac{E_L\beta^4}{8E_C}, \lambda = \frac{E_J\beta^2}{8E_C}$, and $\theta=\Phi_{\mathrm{ext}}-\pi$, where $\Phi_{\mathrm{ext}}$ is a tunable external flux threading each loop, and $\beta \neq 0$ can be chosen arbitrarily (see SM~\cite{supp}). In fact, when $\beta\neq\sqrt{4\pi}$, the circuit describes a more general model known as the massive Thirring-Schwinger model~\cite{frohlichMassiveThirringSchwingerModel1976}. 
	In the SM \cite{supp}, we present a method for preparing initial wave packets of bosonic particles using two ancillary qubits, hence providing a complete protocol for preparation and evolution of mesonic wave packets for a scattering experiment. 
	Measurements of the local density \cite{zhangSuperconductingQuantumSimulator2023} or the output field at the edges \cite{forn-diazUltrastrongCouplingSingle2017,vrajitoareaUltrastrongLightmatterInteraction2022a} can be performed using standard techniques.
	
	{\it Discussion and outlook.---}
	Using \emph{ab initio} numerical uMPS computations and working with a bosonized formulation of the Schwinger model, we analyzed the real-time dynamics of high-energy particle scattering in the nonperturbative regime of QED in 1+1 dimensions. We also proposed an analog circuit-QED implementation of the bosonized Schwinger model. This implementation  
	requires minimal ingredients and no approximations (besides a lattice discretization), in contrast to previous circuit-QED proposals based on a quantum-link model~\cite{marcosSuperconductingCircuitsQuantum2013a}.
	We studied both the confined and deconfined regimes of the model, exhibiting  
	a multitude of phenomena, including  inelastic particle production, meson disintegration, and dynamical string formation and  breaking. 
	
	The single-particle excitations allowed us to obtain complete time-resolved momentum and position distributions of the outgoing $2\rightarrow 2$ scattered particles.
	To account for higher-order scattering it appears necessary to include states where two particles can be close, which could potentially be accomplished  using the two-particle uMPS ansatz from Ref.~\cite{vanderstraetenMatrixMatrixProduct2014}. This might also shed light on the nontrivial transient dynamics in~\cref{fig:meson_scat}(d).
	It would also be interesting to explore the string dynamics more systematically ~\cite{verdelDynamicalLocalizationTransition2023}.
	Additionally, it will be valuable to investigate how other nonequilibrium phenomena, such as dynamical quantum phase transitions (DQPTs) \cite{zacheDynamicalTopologicalTransitions2019,muellerQuantumComputationDynamical2022}, weak ergodicity breaking, and quantum many-body scars~\cite{Desaules:2022ibp,Desaules:2022kse}, can be probed in scattering processes.

	In addition to exploring scattering in regimes inaccessible to the MPS methods, our circuit-QED implementation can also be used to study quench dynamics. For example, the Schwinger mechanism or DQPTs can be studied in quenches of the $\theta$ parameter~\cite{zacheDynamicalTopologicalTransitions2019,muellerQuantumComputationDynamical2022}, which can be accomplished using time-dependent flux control~\cite{blaisCircuitQuantumElectrodynamics2021}.

	Finally, our circuit-QED implementation applies to other bosonic theories \cite{tianAnalogCosmologicalParticle2017,garcia-alvarezFermionFermionScatteringQuantum2015,royQuantumSineGordonModel2021,roySolitonConfinementQuantum2023}, including the $\phi^4$ theory (achieved in the $\beta\rightarrow 0$ limit) in 1+1 or 2+1 dimensions \cite{supp} and generalizations of the bosonized Schwinger model, including to multi-flavor fermions \cite{colemanMoreMassiveSchwinger1976,banulsDensityInducedPhase2017} and to  Thirring interactions \cite{frohlichMassiveThirringSchwingerModel1976}.
	In the latter case, sufficiently strong Thirring interactions give rise to attractive short-range interactions between quarks in the deconfined phase, as shown in the SM~\cite{supp}, leading to stable meson particles and hence qualitatively different scattering dynamics.

	\begin{acknowledgments}
		{\it Acknowledgments.---}We acknowledge valuable discussion with A.~Milsted and Z.~Minev.
		The uMPS simulations were performed with the help of the MPSKit.jl Julia package (\url{https://github.com/maartenvd/MPSKit.jl}). We thank M.~Van Damme for help with the package. 
		The authors acknowledge the University of Maryland's supercomputing resources (\url{http://hpcc.umd.edu}) made available for conducting the research reported in this paper. 
		R.B., S.W., A.F., and A.V.G.~were supported in part by the National Science Foundation (NSF) Quantum Leap Challenge Institute (award no.~OMA-2120757), Department of Energy (DOE), Office of Science, Office of Advanced Scientific Computing Research (ASCR), Accelerated Research in Quantum Computing program (award no.~DE-SC0020312), ARO MURI, the DOE ASCR Quantum Testbed Pathfinder program (awards no.~DE-SC0019040 and no.~DE-SC0024220),  NSF Physics Frontier Center Quantum Computing program, AFOSR, AFOSR MURI, and DARPA SAVaNT ADVENT. Support is also acknowledged from the DOE, Office of Science, National Quantum Information Science Research Centers, Quantum Systems Accelerator.  	
		N.M.~acknowledges funding by the U.S. Department of Energy, Office of Science, Office of Nuclear Physics, InQubator for Quantum Simulation (IQuS) (\url{https:// iqus.uw.edu}) under Award Number DOE (NP) Award DE-SC0020970 via the program on Quantum Horizons: QIS Research and Innovation for Nuclear Science.  Z.D.~and N.M.~acknowledge funding by the DOE, Office of Science, Office of Nuclear Physics via the program on Quantum Horizons: QIS Research and Innovation for Nuclear Science (award no.~DE-SC0021143). Z.D. further acknowledges support by the DOE, Office of Science, Early Career Award (award no.~DESC0020271). 
		E.R.B acknowledges support from the DOE, Office of Science, Office of ASCR, Computational Science Graduate Fellowship (award no.~DE-SC0023112).
	\end{acknowledgments}

\end{document}


\title{Supplemental Material for:
\\
High-Energy Collision of Quarks and Mesons in the Schwinger Model:
\\
From Tensor Networks to Circuit QED
}
\date{\today}
\author{Ron~Belyansky}
\author{Seth~Whitsitt}
\affiliation{Joint Center for Quantum Information and Computer Science, NIST/University of Maryland, College Park, MD 20742 USA}
\affiliation{Joint Quantum Institute, NIST/University of Maryland, College Park, MD 20742 USA}
\author{Niklas~Mueller}
\affiliation{InQubator for Quantum Simulation (IQuS), Department of Physics, University of Washington, Seattle, WA 98195, USA}
\author{Ali~Fahimniya}
\author{Elizabeth~Bennewitz}
\affiliation{Joint Center for Quantum Information and Computer Science, NIST/University of Maryland, College Park, MD 20742 USA}
\affiliation{Joint Quantum Institute, NIST/University of Maryland, College Park, MD 20742 USA}
\author{Zohreh~Davoudi}
\affiliation{Maryland Center for Fundamental Physics and Department of Physics, University of Maryland, College Park, MD 20742 USA}
\affiliation{Joint Center for Quantum Information and Computer Science, NIST/University of Maryland, College Park, MD 20742 USA}
\author{Alexey~V.~Gorshkov}
\affiliation{Joint Center for Quantum Information and Computer Science, NIST/University of Maryland, College Park, MD 20742 USA}
\affiliation{Joint Quantum Institute, NIST/University of Maryland, College Park, MD 20742 USA}
\maketitle
\onecolumngrid

\renewcommand{\theequation}{S\arabic{equation}}

\renewcommand{\bibnumfmt}[1]{[S#1]}
\renewcommand{\citenumfont}[1]{S#1} 

\pagenumbering{arabic}

\makeatletter
\renewcommand{\thefigure}{S\@arabic\c@figure}
\renewcommand \thetable{S\@arabic\c@table}

This Supplemental Material is organized as follows. In \cref{sec:thirring-schwinger}, we discuss the Thirring-Schwinger model, a generalization of the Schwinger model studied in the main text. In particular, we derive in perturbation theory the effective quark-antiquark Hamiltonian [Eq.~(4) of the main text] and discuss the existence (or lack thereof) of bound states. In \cref{sec:circuit}, we derive a corresponding circuit-QED Hamiltonian (including a generalization to 2 dimensions) and discuss an experimental scheme for wave-packet preparation. In \cref{sec:numerics}, we present the details of the numerical tensor-network methods employed, including uniform matrix product states, wave-packet preparation, and particle detection.

\tableofcontents

\section{The Massive Thirring-Schwinger model}
\label{sec:thirring-schwinger}
\noindent
In this section, we discuss the massive Thirring-Schwinger model, realized in our circuit-QED proposal, including its bosonization, Hamiltonian formulation, and the presence of quark-antiquark bound states for different parameters. 
\subsection{Hamiltonian and bosonic dual}
\label{sec:bosonization}%
Consider the Lagrangian density for the so-called massive Thirring-Schwinger model 
\begin{equation}
\label{eq:L-Thirring-Schwinger}
	\mathcal{L} =  \bar{\psi} (i\slashed{\partial}-e\slashed{A}-m)\psi -\frac{1}{4}F_{\mu\nu}F^{\mu\nu}- \frac{1}{2}g(\bar{\psi}\gamma^\mu\psi)(\bar{\psi}\gamma_\mu\psi)\,,
\end{equation}
which for $g=0$ reduces to the massive Schwinger model investigated in this study [Eq.~(1) in the main text], while $e=0$ yields the massive Thirring model~\cite{frohlichMassiveThirringSchwingerModel1976,colemanQuantumSineGordonEquation1975}.
The gauge fields can
be eliminated using Gauss's law~\cite{colemanMoreMassiveSchwinger1976}, which, after fixing the gauge to $A_1(x)=0$, reads
\begin{equation}
	\label{eq:Gauss}
	\partial_x E = - \partial_x^2 A_0 = e\rho,
\end{equation}
where $\rho(x) = \psi^\dagger(x) \psi(x)$ is the charge-density operator. 
The solution to this equation is
\beq
\label{eq:A0}
A_0(x) = a_0 - \frac{e \theta}{2 \pi} x - \frac{e}{2} \int dx' \, \rho(x') |x - x'| ,
\eeq
where $a_0$ and $\theta$ are integration constants. As argued in Ref.~\cite{colemanMoreMassiveSchwinger1976}, physics depends on $\theta$ only modulo $2\pi$, and so a suitable range for this variable is $\theta \in (-\pi,\pi]$. The Hamiltonian can be derived in the standard fashion, noting the expression for the electric field from \cref{eq:Gauss} with $A_0$ given in \cref{eq:A0}.
In the charge-zero subspace $\int dx \, \rho(x) = 0$, the (normal-ordered) Hamiltonian of the Thirring-Schwinger model becomes 
\begin{align}
H =& \int dx \, :\psi^{\dagger} \gamma^0 \left( - i \gamma^1 \partial_x + m \right) \psi: - \frac{e^2 \theta}{2 \pi} \int dx \, x \, :\rho(x): \, \nn
& - \frac{e^2}{4} \int dx \int dx' \, |x - x'|  :\rho(x) \rho(x'): + \frac{g}{2} \int dx :\left[ \rho(x)^2 - \left( \psi^{\dagger}(x) \gamma^0 \gamma^1 \psi(x) \right)^2 \right]\,.
\label{eq:TSH}
\end{align}
Our conventions are such that $\{ \gamma^{\mu} , \gamma^{\nu} \} = 2 \eta^{\mu \nu}$, with metric signature $\eta^{00} = -\eta^{11} = +1$.
The two-component spinor operators, $\psi(x,t)\equiv (\psi_1(x,t),\psi_2(x,t))^T$, satisfy the canonical anticommutation relations $\{ \psi_a(x,t),\psi^{\dagger}_b(x',t) \} = \delta_{ab} \delta(x - x')$, where $a,b = 1,2$. 
The model described above can be shown to be dual to a bosonic theory with the Hamiltonian~\cite{colemanChargeShieldingQuark1975,colemanMoreMassiveSchwinger1976,frohlichMassiveThirringSchwingerModel1976,colemanQuantumSineGordonEquation1975}
\begin{equation}
	\label{eq:Schwinger-Ham-before-shift}
	H =  \int dx \, :\left[\frac{\Pi^2}{2} +\frac{(\partial_x\phi)^2}{2}+\frac{M^2(\phi+\theta/\beta)^2}{2}-u\cos(\beta\phi)\right]:\,,
\end{equation}
where $[\phi(x),\Pi(y)]=i\delta(x-y)$ and $[\Pi(x),\Pi(y)]=[\phi(x),\phi(y)]=0$ and the normal-ordering is defined with respect to the boson mass $M$ \cite{colemanMoreMassiveSchwinger1976,colemanChargeShieldingQuark1975}.
The model parameters are related to those in the fermionic model as follows: 
\begin{equation}
	\label{eq:thirring-schwinger-params}
	M = \frac{e}{\sqrt{\pi}}\frac{1}{\sqrt{1+{g}/{\pi}}},\quad u =\frac{\exp(\gamma)}{2\pi^{3/2}}m\,e, \quad \beta = \sqrt{\frac{4\pi}{1+{g}/{\pi}}}\,,
\end{equation}
with $\gamma$ being the Euler's constant.
Furthermore, the following relation holds between the fermionic current $\overline{\psi} \gamma^\mu \psi$ and the bosonic field $\phi$~\cite{colemanQuantumSineGordonEquation1975}:
\begin{equation}
\label{eq:current-conversions}
        \overline{\psi} \gamma^\mu \psi \, = \, -\frac{\beta}{2\pi}\epsilon^{\mu \nu} \partial_\nu \phi.
\end{equation}
Here, $\epsilon^{\mu \nu}$ is the Levi-Civita tensor.
Now, using $\partial_x E = e \overline{\psi}\gamma^0\psi$ [see \cref{eq:Gauss}], one arrives at $\frac{e\beta}{2\pi}\phi = E$, which relates the scalar field $\phi$ to the electric field $E$. 
It is more convenient to work with a shifted $\phi$: $\phi\rightarrow \phi-\theta/\beta$, such that the Hamiltonian is
\begin{equation}
	\label{eq:Schwinger-Ham}
	H =  \int dx \, :\left[\frac{\Pi^2}{2} +\frac{(\partial_x\phi)^2}{2}+\frac{M^2\phi^2}{2}-u\cos(\beta\phi-\theta)\right]:\,,
\end{equation}
and the relation to the electric field is now $\frac{e\beta}{2\pi}\phi = E +\frac{e\theta}{2\pi}\equiv E_T$, which in the limit $g=0$ reproduces the relation presented in the main text between the total electric field $E_T$ and the bosonic field of the massive Schwinger model. 
When $g=0$, \cref{eq:Schwinger-Ham} reduces to Eq.~(2) of the main text. 
 
\subsection{Quark-antiquark interactions and bound states}
\label{sec:quark-potential}
In this section, we derive an effective quark-antiquark Hamiltonian in the nonrelativistic limit in perturbation theory and use this Hamiltonian to confirm the existence of quark-antiquark bound states (mesons). We also study the meson bound states using  nonperturbative tensor-network computation of the low-lying spectrum.

\subsubsection{Derivation of effective  Hamiltonian}
The goal is to derive an effective interaction Hamiltonian between a quark and an antiquark to leading order in the interactions $e$ and $g$. For $g=0$, an analogous computation was discussed by Coleman in Ref.~\cite{colemanMoreMassiveSchwinger1976}.
The idea is that, in the weak-coupling limit in which $e/m,g \ll 1$, one can restrict the physics to subspaces with fixed particle number, e.g., vacuum, quark-antiquark state, etc.~\footnote{Recall that we have restricted the model to the net zero electric-charge sector.}, since transitions between states with different particle numbers are higher order in the coupling strength~\cite{colemanMoreMassiveSchwinger1976}. This means that, in this limit, the full Hamiltonian can be assumed to be almost block-diagonal in the Fock basis. This mimics a nonrelativistic limit in which one can define an `\emph{effective}' potential describing interactions in each fixed-particle sector, with well-defined quantum-mechanical operators $\hat{x}$ and $\hat{p}$ which would have been meaningless otherwise. Relativistic corrections in the dynamics can be included using standard quantum-mechanical perturbation theory, while corrections in the kinematics can be included by incorporating higher-order terms in $p/m$, where $p$ is the typical momentum in the system. Such a notion of an effective Hamiltonian is useful to get a qualitative understanding of the nature of quark-antiparticle interactions in such a limit, and to make analytic predictions for the expected spectrum that can be compared against exact numerics.

To start with, we keep the kinematics relativistic but constrain our analysis to the two-particle sector only. A nonrelativistic expansion in $p/m$ will be performed at the end. First, note that the quadratic piece of the Hamiltonian in \cref{eq:TSH}, $H_0\equiv  \int dx \, :\psi^{\dagger} \gamma^0 \left( - i \gamma^1 \partial_x + m \right) \psi:$, can be diagonalized by a standard mode expansion:
\beq
\psi_a(x) = \int \frac{dk}{\sqrt{4 \pi E_k}} e^{i k x} \left[ u_a(k) b(k) + v_a(-k) c^{\dagger}(-k) \right],
\label{eq:modeexp}
\eeq
where $E_k \equiv \sqrt{k^2 + m^2}$ with $k\equiv k^1=-k_1$, and $u(k)$ and $v(k)$ are two-component spinor wave functions that  satisfy the classical Dirac equation for positive and negative frequencies, respectively: $(\slashed{k} - m) u(k) = (\slashed{k} + m)v(k) = 0$ (here, $\slashed{k}=E_k \gamma^0-k\gamma^1$). The creation operators for quarks and antiquarks satisfy the canonical anticommutation relations
$\{ b(k),b^{\dagger}(k') \} = \{ c(k),c^{\dagger}(k') \} = \delta(k - k')$.
Further, the following representation of Dirac matrices is used for  explicit computations, 
$\gamma^0 = \sigma^z$ and 
$ \gamma^1 = i \sigma^y$,
so that the spinor wavefunctions are given by
$u(k) = \begin{pmatrix}
\sqrt{m + E_k}
\\
\frac{k}{\sqrt{m + E_k}}
\end{pmatrix}$ and 
$v(k) = \begin{pmatrix}
\frac{k}{\sqrt{m + E_k}} \\
\sqrt{m + E_k}
\end{pmatrix}$,
leading to the free-fermion Hamiltonian
\beq
H_0 = \int dk \, E_k \left[ b^{\dagger}(k) b(k) + c^{\dagger}(k) c(k) \right].
\eeq
A quark-antiquark state in the noninteracting limit can be written as
\beq
| p, q \rangle = b^{\dagger}(p) c^{\dagger}(q) | 0 \rangle,
\eeq
where $| 0 \rangle$ is the Fock vaccum of $H_0$ and where the quark (antiquark) has momentum $p$ ($q$).
Following Coleman~\cite{colemanMoreMassiveSchwinger1976}, one can now \emph{define} the reduced center-of-mass Hamiltonian as the operator $H_{\mathrm{eff}}$, whose matrix elements in the two-particle sector are given by
\beq
\langle p', q' | H | p, -p \rangle = \delta(p' + q') \langle p' | H_{\mathrm{eff}} | p \rangle.
\eeq
The effective Hamiltonian $H_{\mathrm{eff}}$ is a function of a conjugate pair of operators $[\hat{x},\hat{p}] = i$, where $\hat{x}$ is the displacement between the quark and the antiquark and  $|p\rangle$ is a single-particle momentum eigenstate, $\hat{p}|p\rangle = p |p \rangle$, with normalization $\langle p'|p\rangle = \delta (p - p')$.
In the absence of interactions, one has
\bea
\langle p', q' | H_0 | p, -p \rangle &= 2 \sqrt{m^2  + p^2} \, \delta(p - p') \delta(p' + q') \quad
\Rightarrow \quad H_{\mathrm{eff}} = 2 \sqrt{m^2 + \hat{p}^2 }\,.
\eea
To compute $\langle p', q' | H | p, -p \rangle$ in the interacting case, one can insert \cref{eq:modeexp} into \cref{eq:TSH} to obtain
\begin{align}
\langle p' | H_{\mathrm{eff}} | p \rangle &= 2 \sqrt{m^2  + p^2} \, \delta(p - p') - \frac{e^2 \theta}{4 \pi^2} \int dx \, x \, e^{i (p - p')x}  +  \frac{e^2}{8 \pi E_{p'} E_p} \left( m^2 + p' p + E_{p'} E_p \right)\int dx \, |x| \, e^{i (p - p')x} \nn & + \frac{e^2}{8 \pi} \frac{m^2}{E_{p'}^2 E_p^2} - \frac{g}{2 \pi E_{p'} E_p}\left( m^2 + p' p + E_{p'} E_p \right).
\label{eq:HeffP}
\end{align}
Note that all interaction terms are of the form
\beq
\langle p' | \hat{O} | p \rangle = f_1(p') f_2(p) \int \frac{dx}{2 \pi} \, g(x) \, e^{i (p - p')x}
\label{eq:Hform}
\eeq
for some functions $f_1$, $f_2$, and $g$ [for the last line of Eq.~\eqref{eq:HeffP}, $g(x) = \delta(x)$].
This allows any interaction term $\hat{O}$ to be written as follows:
\begin{align}
\hat{O} &= \left(\int dp' \, |p' \rangle \langle p' | \right)  \hat{O} \left( \int dp \, |p \rangle \langle p | \right) \nn
&= \int dp' \int dp \, | p' \rangle f_1(p') f_2(p) \int \frac{dx}{2 \pi} \, g(x) \, e^{i (p - p')x} \langle p | \nn
&= f_1(\hat{p}) \int dp' \int dp \, | p' \rangle \left( \int dx \, g(x) \langle p' | x \rangle \langle x | p \rangle \right) \langle p | f_2(\hat{p}) \nn 
&= f_1(\hat{p}) \left(\int dp' \, |p' \rangle \langle p' | \right) g(\hat{x}) \left(\int dx \, |x \rangle \langle x | \right) \left(\int dp \, |p \rangle \langle p | \right) f_2(\hat{p}) \nn
&= f_1(\hat{p}) g(\hat{x}) f_2(\hat{p})\,.
\end{align}
\Cref{eq:Hform} is used in the second equality, and the identity $e^{i (p - p')x} = 2 \pi \langle p' | x \rangle \langle x | p \rangle$ is used in the third equality.
Finally, the effective Hamiltonian can be written as
\begin{align}
H_{\mathrm{eff}} &= 2 \left(m^2 + \hat{p}^2 \right)^{1/2} - \frac{e^2 \theta}{2 \pi} \hat{x} + \frac{e^2}{4} \frac{m}{m^2 + \hat{p}^2} \delta(\hat{x}) \frac{m}{m^2 + \hat{p}^2} \nn
& +  \frac{e^2}{4} \left[ | \hat{x} | + \frac{\hat{p}}{\left(m^2 + \hat{p}^2 \right)^{1/2}} |\hat{x}| \frac{\hat{p}}{\left(m^2 + \hat{p}^2 \right)^{1/2}} + \frac{m}{\left(m^2 + \hat{p}^2 \right)^{1/2}} |\hat{x}| \frac{m}{\left(m^2 + \hat{p}^2 \right)^{1/2}} \right] \nn
& - g  \left[ \delta(\hat{x}) + \frac{\hat{p}}{\left(m^2 + \hat{p}^2 \right)^{1/2}} \delta(\hat{x}) \frac{\hat{p}}{\left(m^2 + \hat{p}^2 \right)^{1/2}} + \frac{m}{\left(m^2 + \hat{p}^2 \right)^{1/2}} \delta(\hat{x}) \frac{m}{\left(m^2 + \hat{p}^2 \right)^{1/2}} \right]\,. \label{eq:H-eff-rel-kinematics}
\end{align}

The lengthy expression in \cref{eq:H-eff-rel-kinematics} can be simplified by considering the nonrelativistic limit.
Note that, when momentum and energy are large enough for particle creation, $\langle \hat{p}^2 \rangle \gtrsim m^2$, non-particle-conserving, i.e., inelastic, transitions can occur, and, in such a regime, it is not particularly useful to consider an effective potential between a quark and an antiquark.
However, since our interest is in an effective interaction between static or slow-moving quark and antiquark---in particular, for investigating the presence of bound states---the matrix elements between states with different particle numbers will be reduced by kinematic constraints. Note that this limit is only applicable when the dimensionless coupling constants $e/m$ and $g$ are small enough, since large couplings result in binding- or scattering-energy scales large enough to violate these assumptions.
Based on this discussion, $H_{\mathrm{eff}}$ can be expanded in $p/m \ll 1$ to  obtain a simpler effective Hamiltonian at leading order in $p/m$, $e/m$, and $g$:
\beq
H_{\mathrm{eff}} = 2 m + \frac{\hat{p}^2}{m} + \frac{e^2}{2} \left( |\hat{x}| - \frac{\theta}{\pi} \hat{x} \right) + \frac{e^2}{4 m^2} \delta(\hat{x}) - 2 g \delta(\hat{x}).
\label{eq:H_eff_simplified}
\eeq
In taking the large-$m$ limit, the dimensionless combination $e/m$ is kept fixed.
For the Schwinger model, $g=0$, this gives Eq.~(4) in the main text. Note  that the electric ($e\neq 0$) and the Thirring ($g\neq 0$) interactions contribute short-range terms which compete with each other. 
In the confined phase ($\theta\neq \pi$), the linear potential guarantees quark-antiquark bound states, which are the fundamental excitations, regardless of the short-range interactions. However, in the deconfined phase ($\theta=\pi$), quarks are free particles (as long as $x>0$, i.e., the quark is to the left of the antiquark). Here, the presence of bound states depends on the delta-function term in \cref{eq:H_eff_simplified}.
When $g> g_c \equiv \frac{e^2}{8m^2}$, the delta-function term is negative, giving rise to attractive short-range interactions, implying the existence of at least one bound state. For $g< g_c$ (including the $g=0$ case considered in the main text), on the other hand, the delta function is repulsive, prohibiting any bound states from forming.

As a nontrivial check on this expression, consider the Thirring model, $e=0$, which is an integrable quantum field theory whose spectrum is known exactly~\cite{Dashen:1975hd,Zamolodchikov:1978xm}.
The effective Hamiltonian in this case is simply
\beq
H_{\mathrm{eff}} = 2 m + \frac{\hat{p}^2}{m} - 2 g \delta(\hat{x})\,.
\eeq
This is a standard problem in introductory quantum mechanics (see e.g., Ref.~\cite{griffithsquantum}).
For $g>0$, there is a single bound state with energy
\beq
E_{\mathrm{bound}} = 2m \big(1 - \frac{g^2}{2}\big).
\label{eq:approxTM}
\eeq
The exact Thirring model with $g>0$ has $N$ bound states, where $N$ is the largest integer smaller than $1 + 2g/\pi$, and the energy of the $n$th bound state is given by~\cite{Dashen:1975hd}
\beq
E_n = 2m \sin \left( \frac{n \pi/2}{1 + 2g/\pi} \right), \quad n = 1, 2,.., N\,.
\eeq
For small $g$, there is a single bound state ($n=1$) with energy given, at leading order in $g$, by \cref{eq:approxTM}.

\subsubsection{Numerical verification}
\begin{figure}[b!]
	\centering	\includegraphics{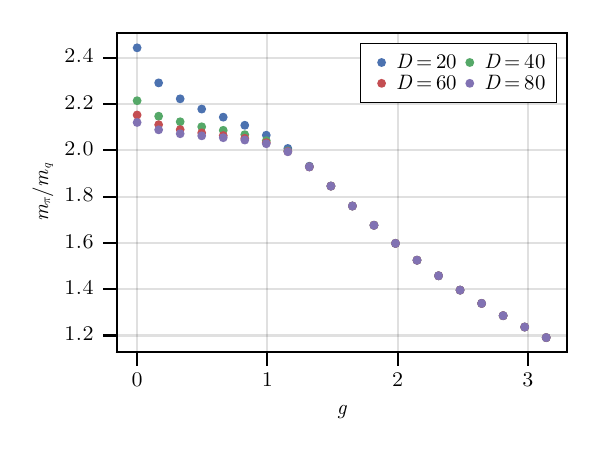}
	\caption{Ratio of the meson mass $m_\pi$ to the quark mass $m_q$ in the deconfined phase ($\theta=\pi$) as a function of the Thirring interaction strength $g$, for different bond dimensions $D$. The energies are calculated using the variational principle with the uMPS quasiparticle ansatz [see \cref{eq:qp-ansatz}] using the same parameters as in the main text ($\chi=1,\mu^2=0.5$, and $\lambda=0.1$).}
	\label{fig:meson_masses}
\end{figure}

To go beyond the perturbative results, we make use of the variational uMPS quasiparticle ansatz [see \cref{sec:uMPS} and \cref{eq:qp-ansatz}] to verify the existence of the bound states. In the deconfined phase, quarks are topological ``kinks"~\cite{colemanMoreMassiveSchwinger1976}, and are numerically described by the topological uMPS ansatz, whereas the mesons, if they exist, would be described by the topologically trivial uMPS ansatz (see \cref{sec:numerics}). 
The energy minimization of the topological uMPS ansatz yields the quark mass $m_q$, and the minimization of the topologically trivial uMPS ansatz returns an energy which we denote $m_\pi$. To determine if this corresponds to a meson eigenstate, we plot the ratio $m_\pi /m_q$ in \cref{fig:meson_masses} as a function of $g$. If the meson exists, that ratio needs to satisfy $m_\pi/ m_q < 2$, since the bound-state energy must be below the two-particle continuum beginning at $2m_q$. Furthermore, plotting this ratio for different values of the bond dimension $D$ of the uMPS ansatz can signify the existence or absence of the meson. This is because, if the meson exists, its wavefunction will be localized, and so the ansatz energy $m_\pi$ should be rather insensitive to $D$ and will quickly converge to the true meson mass as $D$ is increased~\cite{haegemanElementaryExcitationsGapped2013}.

\Cref{fig:meson_masses} reveals a critical $g_c\approx 1$ (for the parameters used in the main text) above which $m_\pi / m_q < 2$. This region clearly shows insensitivity to $D$, signaling that the ansatz properly captures the nature of the bound-state wave function even for smaller bond dimensions. Below $g_c$, on the other hand, $m_\pi /m_q > 2$, and a qualitatively different behavior is observed as a function of $D$, signaling sensitivity to the choice of bond dimension.
All this suggests that the bound state in the deconfined phase only exists for sufficiently large $g$, in agreement with the analytical prediction in the previous section.
In the absence of a bound state excitation, the minimization of the topologically trivial uMPS ansatz can only yield a multi-particle state that is not an eigenstate, but is rather a superposition of many eigenstates from the multi-particle continuum.  
Increasing the bond dimension of such a state leads to a reorganization of the particles in it and a reduction of their interaction energy, 
giving rise to a strong dependence of the state energy on the bond dimension, as can be seen in \cref{fig:meson_masses}.

\section{Circuit-QED implementation}
\label{sec:circuit}
In this section, we derive the circuit-QED Hamiltonian from a lumped-element model and present a scheme for preparing meson excitations.
\subsection{Hamiltonian derivation}
\label{sec:circuit-derivation}
\begin{figure}[b!]	
	\centering
	\begin{circuitikz}[scale=0.9,cross/.style={path picture={ 
  \draw[black]
(path picture bounding box.south east) -- (path picture bounding box.north west) (path picture bounding box.south west) -- (path picture bounding box.north east);
}}]]
		\draw[dashed] (0,0) to (-1,0);
		\draw (0,0) to (2,0)
		to [L,l_=$L$,*-,](2,-2)node[ground]{};
		\draw (0,0) to [C,l_=$C$](0,-2);
		\draw (2,0) to (4,0)
		to [barrier,l_=$I_c$,label distance=-10pt](4,-2) to (0,-2);
		\draw (4,0) to [L,l_=$L'$,i>=$$,v^<=$\phi_{L'}^{i-1,i}$] (6,0);
		\draw (6,0) to[short,i<=$$]
		(8,0) to[short,i<=$$] (10,0) 
		to [barrier,i<=$$,v^>=$\phi_J^i$,bipoles/length=1.2cm](10,-2) to[short,i<=$$] (8,-2)node[ground]{}
		to (6,-2) to [C,i<=$$,v>=$\phi_C^i$] (6,0);
		\draw (8,-2) [L,-*,i>=$$,v<=$\phi_L^i$] to (8,0);
		\draw {[anchor=south] (2,0) node {$\phi_{i-1}$}  (8,0) node {$\phi_{i}$} (14,0) node {$\phi_{i+1}$}};
		\draw (10,0) to [L,l_=$L'$,i>=$$,v^<=$\phi_{L'}^{i,i+1}$] (12,0);
		\draw (12,0)
		to (14,0) to (16,0) 
		to [barrier,](16,-2) to (14,-2)node[ground]{}
		to (12,-2) to [C,] (12,0);
		\draw (14,-2) [L,-*] to (14,0);
		\draw[dashed] (16,0) to (17,0);
		\node[] at (15,-0.8) {$\Phi_{\mathrm{ext}}$};
        \node[draw,circle,cross] at  (15,-1.2) {};
	\end{circuitikz}
	\caption{Circuit diagram implementing the massive Thirring-Schwinger model. An external flux  $\Phi_{\mathrm{ext}}$ threads each L-J loop.}
	\label{fig:circuit}
\end{figure}
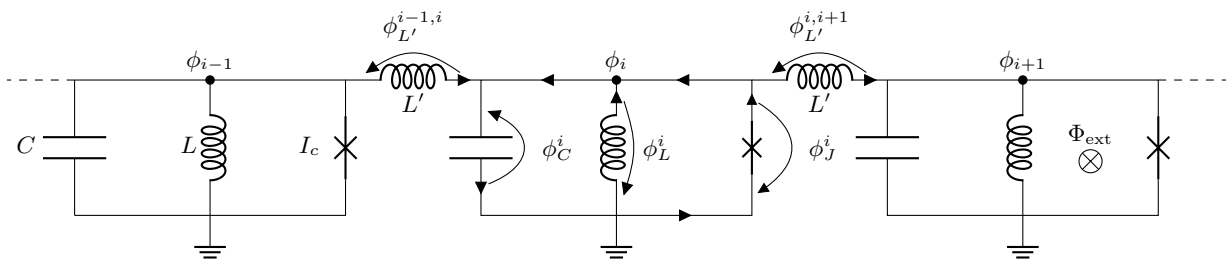

Consider the circuit diagram in \cref{fig:circuit}, which is a more detailed version of Fig.~4 of the main text. Here, each unit cell consists of a capacitor with capacitance $C$, an inductor with inductance $L$, and a Josephson junction with critical current $I_c$, in parallel, representing a general rf-SQUID circuit, which includes the fluxonium as a special case \cite{peruzzoGeometricSuperinductanceQubits2021}. Each L-J loop is threaded by an external flux $\Phi_{\mathrm{ext}}$, and different unit cells are coupled together via inductors with inductance $L'$.
Node fluxes are labeled by $\phi_i$, branch fluxes by $\phi_C^i$, $\phi_J^i$, and $\phi_L^i$, for the corresponding elements within node $i$, and the inter-node branch fluxes coupling nodes $i$ and $i+1$ by $\phi_{L'}^{i,i+i}$.
The branch currents are related to the branch fluxes by $I_C^i =C\ddot{\phi_i}$, $I^i_L=\phi_L^i/L$, $I^{i,i+1}_{L'}=\phi_{L'}^{i,i+1}/L'$, and $I_J^i = I_c\sin(\phi_J^i)$, for the capacitor, inductors, and the Josephson junction, respectively \footnote{We work in units where $\hbar=1$ and the reduced flux quantum $\Phi_0/2\pi = \hbar/(4e^2) = 1$. This $e$ is not to be confused with the $e$ of the Schwinger model.}.

The Hamiltonian of the circuit can be derived by standard means.
The capacitor branch fluxes are chosen to be equal to the node fluxes $\phi_C^i=\phi_i\,\forall\, i$. 
Flux quantization yields the remaining branch fluxes: $\phi_J^i+\phi_C^i=\Phi_{\mathrm{ext}}$, $\phi_C^i+\phi_L^i=0$, $\phi_J^{i-1}+\phi_C^i+\phi_{L'}^{i-1,i}=0$.
Current conservation gives $I_{L'}^{i-1,i}-I_C^i+I_L^i+I_J^i-I_{L'}^{i,i+1}=0$, which, together with the above, yields the equation of motion 
\begin{equation}
	-C\ddot{\phi}_i-\frac{1}{L}\phi_i+I_c\sin(\Phi_{\mathrm{ext}}-\phi_i)+\frac{1}{L'}(\phi_{i-1}-2\phi_i+\phi_{i+1})=0.
\end{equation}
The corresponding Lagrangian is 
\begin{equation}
	\mathcal{L} = \sum_i \left[\frac{C\dot{\phi}_i^2}{2}-\frac{(\phi_{i}-\phi_{i-1})^2}{2L'} -\frac{\phi_i^2}{2L}-I_c\cos(\phi_i-\Phi_{\mathrm{ext}})\right].
\end{equation}
Defining the conjugate momentum $\pi_i = \pdv{\mathcal{L}}{\dot{\phi_i}}=C\dot{\phi}_i$ and imposing the canonical commutation relations $\comm*{\phi_i}{\pi_j}=i\delta_{ij}$, we obtain the Hamiltonian
\begin{equation}
\label{eq:exp-ham}
	H = \sum_i \left[4E_C\pi_i^2+\frac{E_{L'}(\phi_{i}-\phi_{i-1})^2}{2} +\frac{E_L\phi_i^2}{2}+E_J\cos(\phi_i-\Phi_{\mathrm{ext}})\right],
\end{equation}
where we defined the energies
\begin{align}
		E_C = \frac{1}{8C} \qc E_{L'} =\frac{1}{L'}\qc E_L =\frac{1}{L}\qc I_J = E_J.
\end{align}
Redefining $\phi_i\rightarrow \beta\phi_i$ and $\pi_i\rightarrow\pi_i/\beta$, we obtain 
\begin{equation}
\label{eq:exp-ham-supp-like-in-main}
	H = \chi\sum_i \left[\frac{\pi_i^2}{2}+\frac{(\phi_{i}-\phi_{i-1})^2}{2} +\frac{\mu^2\phi_i^2}{2}-\lambda\cos(\beta\phi_i-\theta)\right],
\end{equation}
which is Eq.~(3) of the main text with 
\begin{equation}
	\label{eq:schwinger-experimenta}
\chi = \frac{8E_C}{\beta^2}\qc \frac{E_{L'}\beta^4}{8E_C}=1\qc \mu^2=\frac{E_L\beta^4}{8E_C}\qc \lambda = \frac{E_J\beta^2}{8E_C}\qc \theta=\Phi_{\mathrm{ext}}-\pi.
\end{equation}
Recall that the parameters of the bosonized lattice Hamiltonian are
\begin{equation} \label{eq:params}
\chi = \frac{1}{a}\qc \mu^2=M^2a^2\qc \lambda = ua^2 \exp[2\pi\Delta(a)],
\end{equation}
where 
\begin{equation}
\Delta(a)=\int_{-\pi}^{\pi}\frac{d^2k}{(2\pi)^2}\frac{1}{4\sum_\mu\sin[2](\frac{k_\mu}{2})+(Ma)^2}
\end{equation}
is the Feynman propagator on a lattice evaluated at the origin \cite{ohataMonteCarloStudy2023}. 
Equations (\ref{eq:schwinger-experimenta}) and (\ref{eq:params}) together  constitute a dictionary between the parameters of the  bosonized massive Thirring-Schwinger model and those of the circuit-QED Hamiltonian.

\subsection{Experimental meson--wave-packet preparation}
\label{sec:exp_wavepacket-prep}

In this section, we describe a scheme for preparing initial wave packets, focusing on meson excitations. The quarks, being topological excitations, do not couple to local operators, hence their preparation is left to future work.
Our proposal goes as follows. 
We assume the system [i.e.,~Eq.~(3) of the main text] is cooled down to its ground state in the confined phase. 
We add two ancillary qubits~\cite{jordanQuantumAlgorithmsQuantum2012} far away from each other.
Initializing the qubits in the excited state and coupling them to the system will result in the decay of the two qubit excitations into the system, producing two wave packets of quasiparticles. Choosing a weak coupling will ensure that multi-particle states are not be excited.

To see this, first note that, in terms of the quasiparticle degrees of freedom, \cref{eq:exp-ham-supp-like-in-main} can be re-expressed as follows:
\begin{equation}
    \label{eq:Ham-wavepacketp-prep-bare}
H = \sum_j
\sum_k \omega_{k,j}\Psi_{k,j}^\dagger\Psi_{k,j} + \text{interactions},
\end{equation}
where $k$ is a label for the eigenstates assuming open boundary conditions. $\Psi_{k,j}^\dagger$ and $\Psi_{k,j}$ are the creation and annihilation operators for the $j$th meson with energy $\omega_{k,j}$, i.e.,
\begin{align}
    \Psi_{k,j}^\dagger \ket{\text{vac}} &= \ket{\pi_{k,j}},\\
    \Psi_{k,j} \ket{\text{vac}} &= 0,
\end{align}
where $\ket{\pi_{k,j}}$ are the meson quasiparticles.

Next, consider, for simplicity, a single qubit (e.g., a transmon or a fluxonium \cite{siddiqiEngineeringHighcoherenceSuperconducting2021}) with frequency $\Delta$, coupled at position $i=L$. The addition to \cref{eq:exp-ham-supp-like-in-main} is 
\begin{equation}
\label{eq:qubit-wavepacket-prep-real-space}
H_{\mathrm{qubit}} = \frac{\Delta}{2}\sigma_z + g(t)\sigma_x(a_{L}+a_L^\dagger),
\end{equation}
where $a_i,a_i^\dagger =\frac{\phi_i\pm i\pi_i}{\sqrt{2}}$ are the creation and annihilation operators  satisfying $\comm*{a_i}{a^\dagger_j}=\delta_{ij}$, and $g(t)$ is the coupling (which can be controlled in time using a tunable coupler \cite{rasmussenSuperconductingCircuitCompanionan2021a}).
In terms of mesonic quasiparticles, the entire Hamiltonian [\cref{eq:exp-ham-supp-like-in-main} plus \cref{eq:qubit-wavepacket-prep-real-space}] can be written as 
\begin{equation}
\label{eq:H-wavepacket-prep-RWA-total}
H = \sum_j
\sum_k \omega_{k,j}\Psi_{k,j}^\dagger\Psi_{k,j} +\sum_j
\sum_k \qty[g(t)\lambda_{k,j}\Psi_{k,j}^\dagger \sigma_-  + \mathrm{H.c.}] + \frac{\Delta}{2}\sigma_z,
\end{equation}
under a rotating-wave approximation (RWA) that assumes only a single excitation in the combined qubit-system and hence ignores interactions from \cref{eq:Ham-wavepacketp-prep-bare}. $\lambda_{k,j}\equiv \bra{\pi_{k,j}}a_L+a_L^\dagger\ket{\mathrm{vac}}$ is a matrix element that depends on the overlaps of $a_L$ and $a_L^\dagger$ between the vacuum $\ket{\text{vac}}$ and the meson eigenstates $\ket{\pi_{k,j}}$. 
Turning on $g(t)$, the (initially) excited qubit will decay into the system, producing a mesonic wave packet.
The central momentum and the shape of the resulting wave packet can be controlled by choosing the qubit frequency $\Delta$ and the time dependence of the qubit-system coupling $g(t)$, as described in the Supplementary Methods of Ref.~\cite{yaoTopologicallyProtectedQuantum2013}.
Placing the qubit at the (left) right edge can ensure that only (right-)left-moving excitations are created.
The matrix element $\lambda_{k,j}$ can either be calculated numerically or accounted for (prior to performing the actual scattering experiment) using measurements of the resulting wave-packet shape and a feedback loop.
The wave-packet shape can be determined from, for example, local measurements of the density ($a_x^\dagger a_x$) \cite{zhangSuperconductingQuantumSimulator2023} or from spectroscopic measurements of the transmitted amplitude at the other edge of the system \cite{forn-diazUltrastrongCouplingSingle2017,vrajitoareaUltrastrongLightmatterInteraction2022a}.

\begin{figure}[t!]	
    \centering
\includegraphics{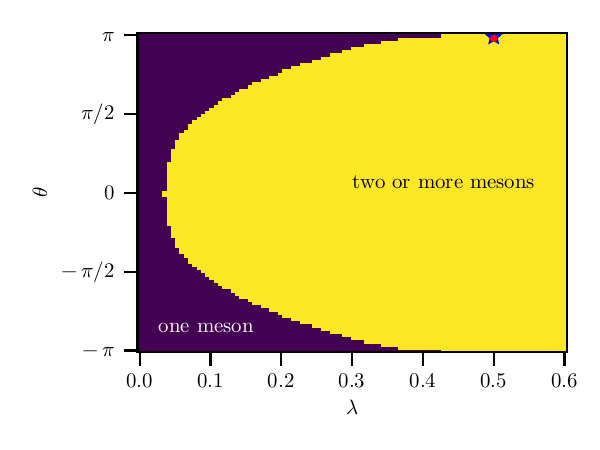}
	\caption{Number of mesons in the confined phase. Purple (yellow) region corresponds to one (two or more) mesons. The two regions are determined by obtaining the two lowest eigenvalues above the ground state using the topologically trivial MPS quasiparticle ansatz (see \cref{sec:uMPS}). 
    The single (two or more) meson region correspond to the second eigenvalue being bigger (smaller) than twice the lowest eigenvalue (mass of the fundamental meson).
The remaining parameters are $\chi=1$ and $\mu^2=0.1$ as in the main text. The star and circle indicate the parameters used in the main text [$\lambda=0.5$ and $\theta=\pi-0.04$ (star) or $\theta=\pi-0.07$ (circle)]. The vertical axis range is $[-\pi+0.001,\pi-0.001]$ so as to avoid the deconfined phase at $\theta=-\pi,\pi$.}
	\label{fig:meson-num-phase-diag}
\end{figure}

An important subtlety is that the qubit generically couples to all mesons in the theory. If there is more than a single meson flavor, this will result in an undesired superposition of wave packets of different mesons. 
To mitigate this issue, this scheme can be combined with adiabatic state preparation. One can first tune the system to a parameter regime where there is only a single meson particle. A simple example is the free-boson limit with $\lambda=0$. More generally, the ``phase diagram" in \cref{fig:meson-num-phase-diag}, obtained using the uMPS methods of \cref{sec:uMPS}, shows the region in the $\{\lambda,\theta\}$ parameter space with only a single meson particle for $\mu^2 = 0.1$.  
This phase diagram is consistent with the perturbative result of Coleman~\cite{colemanMoreMassiveSchwinger1976}, predicting the existence of one meson for $\abs{\theta}\gtrsim \pi/2$ in the limit $\lambda/\mu\rightarrow 0$ (i.e., the strong-coupling limit of the original Schwinger model). 
After preparing the meson wave packets in the single-meson regime, one can adiabatically tune $\lambda$ and $\theta$ to their desired regime, preparing in this way the fundamental mesons of the interacting theory. Tuning both $\theta$ and $\lambda$ can be accomplished using external time-dependant flux control. In order to be able to tune $\lambda$, each Josephson junction in \cref{fig:circuit} can be replaced by a SQUID, a loop composed of two junctions, realizing an effective single flux-tunable junction~\cite{blaisCircuitQuantumElectrodynamics2021}. Designing the two loops (the L-J loop from \cref{fig:circuit} and the SQUID loop) to be asymmetric in size allows one to control both $\lambda(\Phi_\mathrm{ext})$ and $\theta(\Phi_\mathrm{ext})$ with a single external flux~\cite{forn-diazUltrastrongCouplingSingle2017,magazzuProbingStronglyDriven2018,
vrajitoareaUltrastrongLightmatterInteraction2022a}.

\subsection{Generalization to 2+1D and $\phi^4$ models}
\begin{figure}[htb]	
	\centering
	\includegraphics[width=0.95\linewidth]{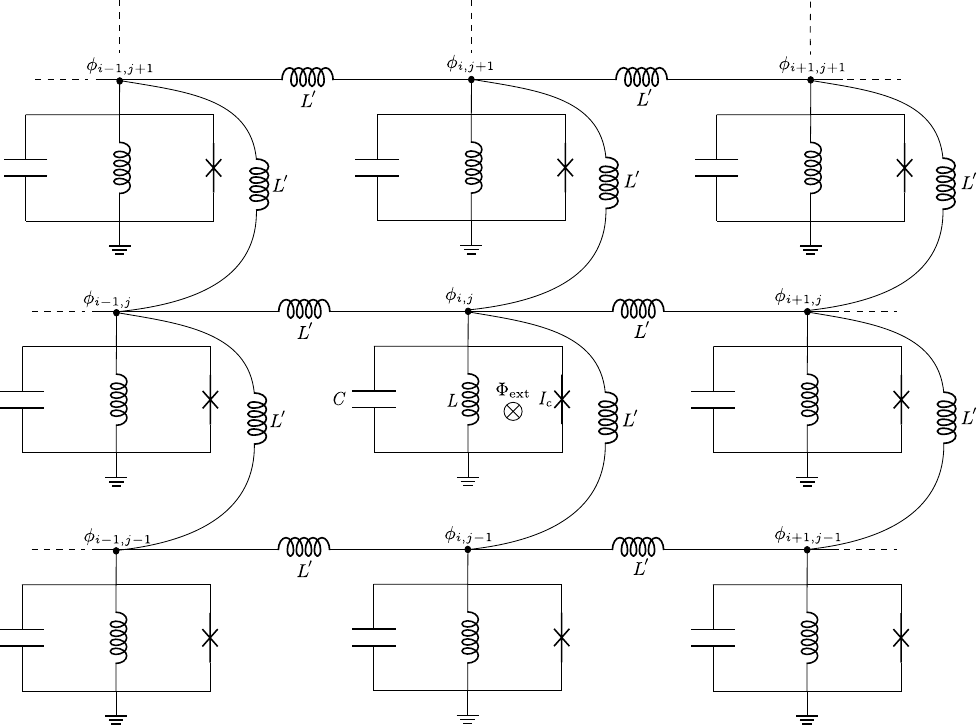}
	\caption{Circuit diagram generalizing \cref{fig:circuit} to two spatial dimensions.}
	\label{fig:circuit2d}
\end{figure}

In this section, we provide a generalization of the circuit from \cref{sec:circuit-derivation} to two spatial dimensions and discuss how it can be used to approximate other field theories such as the $\phi^4$ theory. 
The generaelized circuit, shown in \cref{fig:circuit2d}, consists of identical horizontal layers,  each identical to \cref{fig:circuit}, that are coupled vertically by additional inductors (with the same inductance $L'$ as the horizontal coupling inductors).

A similar derivation to the one described in \cref{sec:circuit-derivation} yields the Hamiltonian
\begin{equation}
    H = \chi\sum_{i,j} \left[\frac{\pi_{i,j}^2}{2}+\frac{(\phi_{i.j}-\phi_{i-1,j})^2+(\phi_{i,j}-\phi_{i,j-1})^2}{2} +\frac{\mu^2\phi_{i,j}^2}{2}-\lambda\cos(\beta\phi_{i,j}-\theta)\right],
\end{equation}
where $i,j$ denote the lattice-site coordinate along the two Cartesian directions, and the parameters are the same as in \cref{eq:schwinger-experimenta}.
This Hamiltonian may be regarded as a lattice discretization of a 2+1D massive Sine Gordon model. While this bosonic model is not dual to 2+1D QED, it can be of interest in its own right.
In addition, this circuit can approximate other theories of interest. In particular, for $\theta=\pi$ and small $\beta$, we obtain an approximation of the lattice-discretized $\phi^4$ model:
\begin{equation}
H\approx  \chi\sum_{i,j} \left[\frac{\pi_{i,j}^2}{2}+\frac{(\phi_{i.j}-\phi_{i-1,j})^2+(\phi_{i,j}-\phi_{i,j-1})^2}{2} +\frac{m^2\phi_{i,j}^2}{2}+\frac{\tilde{\lambda}}{4!}\phi_{i,j}^4\right],
\end{equation}
where $m^2\equiv \mu^2-\lambda\beta^2$ and $\tilde{\lambda}\equiv\lambda \beta^4$.

\section{Numerical Methods}
\label{sec:numerics}
In this section, we provide more details on the uniform-matrix-product-state methods, describing the wave-packet preparation and particle detection.
\subsection{Uniform Matrix Product States}\label{sec:uMPS}
We begin with a general review of 
uniform matrix product states (see Ref.~\cite{vanderstraetenTangentspaceMethodsUniform2019} for more details).
A uniform matrix product state (uMPS), describing a translationally invariant state, is graphically represented as
\begin{equation} 
	\label{eq:uMPS}
	\ket{\Psi(A)} = \cdots
	\begin{diagram}
		\draw (0.5,1.5) -- (1,1.5); 
		\draw[rounded corners] (1,2) rectangle (2,1);
		\draw (1.5,1.5) node (X) {$A$};
		\draw (2,1.5) -- (3,1.5); 
		\draw[rounded corners] (3,2) rectangle (4,1);
		\draw (3.5,1.5) node {$A$};
		\draw (4,1.5) -- (5,1.5);
		\draw[rounded corners] (5,2) rectangle (6,1);
		\draw (5.5,1.5) node {$A$};
		\draw (6,1.5) -- (7,1.5); 
		\draw[rounded corners] (7,2) rectangle (8,1);
		\draw (7.5,1.5) node {$A$};
		\draw (8,1.5) -- (9,1.5); 
		\draw[rounded corners] (9,2) rectangle (10,1);
		\draw (9.5,1.5) node {$A$};
		\draw (10,1.5) -- (10.5,1.5);
		\draw (1.5,1) -- (1.5,.5); \draw (3.5,1) -- (3.5,.5); \draw (5.5,1) -- (5.5,.5);
		\draw (7.5,1) -- (7.5,.5); \draw (9.5,1) -- (9.5,.5); 
		\draw (3.5,0) node {$s_{n-1}$}; \draw (5.5,0) node {$s_{n}$}; \draw (7.5,0) node {$s_{n+1}$};
	\end{diagram}\cdots\,,
\end{equation}
where $A^s$ is a $D\times D$ matrix for each basis index $s$.
When dealing with a bosonic theory, even the local Hilbert space is infinite dimensional and needs to be truncated.
For the parameters used in the main text, we found the local energy basis to be an efficient choice, i.e., the local (single-site) part of the Hamiltonian in Eq.~(3) of the main text was diagonalized using a very large Fock-state basis (of $\sim 2000$ levels), keeping only the lowest $d$ eigenstates (we found $d=12$ to be sufficient for the scattering considered in the main text).
The full Hamiltonian was then projected on this truncated local eigenbasis, and the ground state was subsequently found using variational methods~\cite{zauner-stauberVariationalOptimizationAlgorithms2018}.

The quasiparticle excitations on top of the ground state can be described with the MPS quasiparticle ansatz~\cite{haegemanElementaryExcitationsGapped2013,haegemanVariationalMatrixProduct2012,zauner-stauberVariationalOptimizationAlgorithms2018}
\begin{equation} 
	\label{eq:qp-ansatz}
	\ket{\Phi_p(B)} = \sum_{n} e^{ipn}\;\cdots
	\begin{diagram}
		\draw (0.5,1.5) -- (1,1.5); 
		\draw[rounded corners] (1,2) rectangle (2,1);
		\draw (1.5,1.5) node (X) {$A_L$};
		\draw (2,1.5) -- (3,1.5); 
		\draw[rounded corners] (3,2) rectangle (4,1);
		\draw (3.5,1.5) node {$A_L$};
		\draw (4,1.5) -- (5,1.5);
		\draw[rounded corners] (5,2) rectangle (6,1);
		\draw (5.5,1.5) node {$B$};
		\draw (6,1.5) -- (7,1.5); 
		\draw[rounded corners] (7,2) rectangle (8,1);
		\draw (7.5,1.5) node {$\tilde{A}_R$};
		\draw (8,1.5) -- (9,1.5); 
		\draw[rounded corners] (9,2) rectangle (10,1);
		\draw (9.5,1.5) node {$\tilde{A}_R$};
		\draw (10,1.5) -- (10.5,1.5);
		\draw (1.5,1) -- (1.5,.5); \draw (3.5,1) -- (3.5,.5); \draw (5.5,1) -- (5.5,.5);
		\draw (7.5,1) -- (7.5,.5); \draw (9.5,1) -- (9.5,.5);
		\draw (1.5,0) node {$\dots$}; \draw (3.5,0) node {$s_{n-1}$}; \draw (5.5,0) node {$s_{n}$}; \draw (7.5,0) node {$s_{n+1}$}; \draw (9.5,0) node {$\dots$};
	\end{diagram}\cdots \,.
\end{equation}
This state is written in the so-called mixed canonical form, with the ground-state tensors $A_L$ and $\tilde{A}_R$ in the  left- and right-orthonormal forms, respectively.
$A_L$ and $\tilde{A}_R$ can either represent the same ground state for a topologically trivial excitation, in which case they are related by a gauge transformation (i.e $A_L=C^{-1}A_RC$ for a $D\times D$ matrix $C$), or different degenerate ground states for a topological excitation in case of a spontaneously broken symmetry. The variational optimization of the $B$ tensor reduces to an eigenvalue problem for each $p\in[-\pi,\pi)$, providing both the dispersion relation $\mathcal{E}(p)$ and the $p$-dependent eigenvectors $B(p)$.

\begin{figure}[h!]	\centering
	\includegraphics[width=\linewidth]{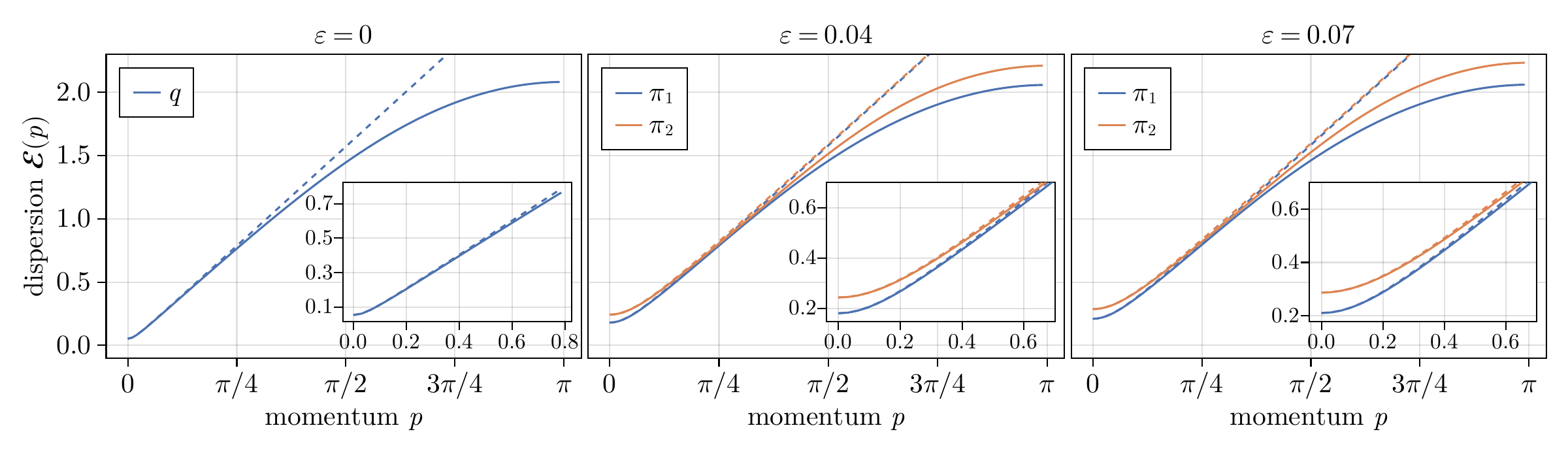}
	\caption{Dispersion relation as a function of (positive) momentum for the three parameter regimes considered in the main text [with $\mathcal{E}(-p)=\mathcal{E}(p)$]. 
Solid lines are the numerical uMPS results [obtained using \cref{eq:qp-ansatz}], and dashed lines are the relativistic approximations $\mathcal{E}(p)\approx\sqrt{m^2+p^2}$, where $m$ is $\mathcal{E}(p=0)$ for the corresponding particle: quarks $q$ in the deconfined phase ($\varepsilon=0$) and mesons $\pi_j$ in the confined phase with $\varepsilon=0.04,0.07$ (only the lightest two mesons corresponding to $j=1,2$ are shown). Insets show a zoom in the low-energy regime.}
	\label{fig:dispersion}
\end{figure}

The dispersion relation is shown in \cref{fig:dispersion} for the three parameter regimes studied in the main text: the deconfined phase where $\theta=\pi$, and the confined phase where $\theta=\pi-\varepsilon$ with $\varepsilon=0.04,0.07$. At low energies (insets), the dispersion is well approximated by the relativistic relation $\mathcal{E}(p)\approx \sqrt{m^2+p^2}$, where $m$ is the mass of the particle [obtained from  $m\equiv \mathcal{E}(p=0)$].

\subsection{MPS wave-packet preparation}\label{sec:wavepackets_prep}
In this section, we describe the numerical procedure for preparing initial wave-packet states using the uMPS quasiparticles. We follow a procedure similar to the one in Refs.~\cite{vandammeRealtimeScatteringInteracting2021,milstedCollisionsFalseVacuumBubble2022} albeit with some differences that are explained below. Using the momentum quasiparticle eigenstates in \cref{eq:qp-ansatz}, a localized wave packets can be built as
\begin{equation}
	\label{eq:wavepacket-from-phik}
\begin{aligned}
	\ket{\Psi_{\mathrm{wp}}} = \int_{-\pi}^{\pi}dp \, c_p\, \ket{\Phi_p\big(B(p)\big)}
	=\sum_n  \cdots
	\begin{diagram}
		\mpsT{-4.5}{0}{A_L} \mpsT{-2.5}{0}{A_L} \mpsT{-0.5}{0}{B_n} \mpsT{1.5}{0}{\tilde{A}_R} \mpsT{3.5}{0}{\tilde{A}_R}
		\draw (-4.5,-1.5) node {$\dots$}; \draw (-2.5,-1.5) node {$s_{n-1}$}; \draw (-0.5,-1.5) node {$s_n$}; 
		\draw (1.5,-1.5) node {$s_{n+1}$}; \draw (3.5,-1.5) node {$\dots$};
		\lineH{-5.5}{-5}{0} \lineH{-4}{-3}{0} \lineH{-2}{-1}{0} \lineH{0}{1}{0} \lineH{2}{3}{0} \lineH{4}{4.5}{0} 
		\draw (-4.5,-.5) -- (-4.5,-1) ; \lineV{-.5}{-1}{-2.5} \lineV{-.5}{-1}{-0.5} \lineV{-.5}{-1}{1.5} \lineV{-.5}{-1}{3.5} 
		\dobase{0}{0}
	\end{diagram} \cdots\,,
\end{aligned}
\end{equation}
where
\begin{equation}
	\label{eq:Bn-def}
B_n \equiv \int_{-\pi}^{\pi}dp\, c_p \, e^{ipn} B(p).
\end{equation}
Here, $c_p\sim e^{-(p-p_0)^2/(2\sigma^2)}$, which is a Gaussian function centered at $p_0$ with width $\sigma$. We use $\sigma=0.12$ for the quark-antiquark wave packets  (Fig.~2 of the main text) and $\sigma=0.06$ for the meson wave packets (Fig.~3 of the main text).

As first discussed in Ref.~\cite{vandammeRealtimeScatteringInteracting2021}, there are several caveats with this approach. 
First, because the $B(p)$ tensors are determined from an eigenvalue problem for each momentum $p$, they come with random phases, preventing the phase coherence needed for building localized wave packets.
To deal with this issue, we fix the global phase of each $B(p)$ such that one specifically chosen tensor element in each is real and positive. 
Second,  one needs to fix the gauge redundancy of the $B(p)$ tensors (i.e., invariance of $\ket{\Phi_p\big(B(p)\big)}$ under $B^s(p)\rightarrow B^s(p)+Y\tilde{A}_R^s-e^{ip}A^s_LY$ for any $D\times D$ matrix $Y$). The choices used when solving the variational minimization problem, the so-called left or right gauge fixing conditions, are inherently very asymmetric. To deal with this issue, we employ the ``reflection symmetric" gauge choice of Ref.~\cite{milstedCollisionsFalseVacuumBubble2022} as it is applicable for both topologically trivial and nontrival excitations.
Together, this approach is simpler and does not require any approximations or conditions on the wave-packets' width, unlike the methods of Refs.~\cite{vandammeRealtimeScatteringInteracting2021,milstedCollisionsFalseVacuumBubble2022}.
The final wave packet ends up slightly shifted from its intended location [which is $n=0$ for the choice of $c_p$ in \cref{eq:Bn-def}], which can be corrected by centering it based on $\mathrm{argmax}{\norm{B_n}}$, as discussed in the following.

Finally, we note that, to evaluate \cref{eq:Bn-def}, one has to sample a finite grid of momenta $p$ with resolution $\Delta p$, and the $B_n$ are, therefore, periodic with period $N_p=\frac{2\pi}{\Delta p}$, i.e., $B_{n+N_p}=B_n$. 
We choose $N_p$ large enough so that the wave packet comfortably fits inside a single period (we used $N_p=400$), which we take to be centered around $n^*=\mathrm{argmax}{\norm{B_n}}$. Of this $N_p$-sized interval of $B_n$ tensors, we only keep $N<N_p$ tensors that satisfy $\norm{B_n}/\norm{B_{n^*}}\ge \epsilon$ for some chosen threshold $\epsilon$ (we used $\epsilon=10^{-3}$), which we label by $n\in[i+1,i+N]$ for a chosen $i$ along the uMPS chain.
The integral in \cref{eq:wavepacket-from-phik} becomes a finite sum and can be analytically expressed as
\begin{equation}
	\label{eq:wavepacket-summed}
\begin{aligned}
	\ket{\Psi_{\mathrm{wp}}}=&\cdots
	\begin{diagram}
		\mpsT{-2.5}{0}{A_L} \mpsTlonger{-0.5}{0}{M_{i+1}} \mpsTlonger{2}{0.0}{M_{i+2}} \mpsTlonger{6}{0}{M_{i+N}} \mpsT{8}{0}{\tilde{A}_R}
		\draw (-2.5,-1.5); \draw (-0.5,-1.5); 
		\draw (1.5,-1.5); \draw (4,0) node {$\dots$};
		\lineH{-3.5}{-3}{0} \lineH{-2}{-1.5}{0} \lineH{0.5}{1}{0} \lineH{3}{3.5}{0} \lineH{4.5}{5}{0} 
		\lineH{7}{7.5}{0} \lineH{8.5}{9}{0}
		\lineV{-.5}{-1}{-2.5} \lineV{-.5}{-1}{-0.5} \lineV{-.5}{-1}{2} \lineV{-.5}{-1}{6}  \lineV{-.5}{-1}{8} 
		\dobase{0}{0}
	\end{diagram} \cdots\,,\\
M_{i+1} =& \mqty(A_L & B_{i+1})\qc M_{n\in [i+2,i+N-1]} = \mqty(A_L & B_n\\0&\tilde{A}_R)\qc M_{i+N} = \mqty(B_{i+N}\\ \tilde{A}_R).
\end{aligned}
\end{equation}

To create the initial two-particle state, this procedure can be carried out twice to create two wave packets with opposite momenta. Given that $A_L$ is related to $A_R$ by a gauge transformation, i.e., $A_LC=CA_R$, the states are then glued together using the matrix $C^{-1}$.
The result is a nonuniform window of tensors surrounded by the uniform ground state.
Note that the bond dimension of the tensors in \cref{eq:wavepacket-summed} is not uniform, as some have the value $D$ and others $2D$. 
Finally, the one-site time-dependent variational principle (TDVP) \cite{Haegeman2016} is  used to evolve the state in time, which does not change the bond dimensions. Hence,  
prior to the time-evolution, we expand the bond dimensions of all the tensors to a uniform value $D' \ge 2D$. For the quark-antiquark scattering (Fig.~2 of the main text), we found that $D=20$ and $D'=50$ were sufficient for convergence at late times, while the meson-meson scattering in the confined phase required a larger bond dimension ($D=40$ and $D'=100$).
During the time evolution, we only update the tensors inside the nonuniform window \cite{milstedVariationalMatrixProduct2013b,phienDynamicalWindowsRealtime2013,zaunerTimeEvolutionComoving2015}. To decide whether to extend the window, at each time step, we compute the entanglement entropy across the bonds at the edges of the window and compare that to the entropy of the vacuum, extending the window by a site if the relative difference is greater than a specified threshold (which we chose to be $0.02$).

\subsection{Particle detection}\label{sec:particle-det}
In this section, we describe how the scattering matrix can be computed by projecting on multi-particle basis states.
As long as the particles are well separated, one can use \cref{eq:qp-ansatz} to build multi-particle states, akin to the asymptotic ``out'' states in the definition of the S-matrix.
For example, a two-particles state, with a particle on the left with momentum $p_1$ and a particle on the right with momentum $p_2$, is expressed as
\begin{align}
	\label{eq:two-particle-basis}
	\ket{\Phi_{p_1,p_2}}=\sum_{n_1\in{W_L}}\sum_{n_2\in{W_R}}e^{ik_1n_1}e^{ik_2n_2}\cdots
	\begin{diagram}
		\mpsT{-4.5}{0}{A_L}  \mpsTlonger{-2.5}{0}{B(p_1)} \mpsT{-0.5}{0}{\tilde{A}_R} \mpsTlonger{3}{0}{C^{-1}} \mpsT{5}{0}{\tilde{A}_L} \mpsTlonger{7}{0}{B(p_2)} \mpsT{9}{0}{A_R}
		\draw (-2.5,-1.5); \draw (-0.5,-1.0); 
		\draw (3.5,-1.5); \draw (1,0) node {$\dots$};
		\lineH{-5.5}{-5}{0} \lineH{-4}{-3.5}{0} \lineH{-1.5}{-1}{0} \lineH{0}{0.5}{0} \lineH{1.5}{2}{0} \lineH{4}{4.5}{0} 
		\lineH{5.5}{6}{0} \lineH{8}{8.5}{0} \lineH{9.5}{10}{0} 
		\lineV{-.5}{-1}{-4.5} \lineV{-.5}{-1}{-2.5} \lineV{-.5}{-1}{-0.5} \lineV{-.5}{-1}{5}  \lineV{-.5}{-1}{7} \lineV{-.5}{-1}{9}
		\draw (-2.5,-1.5) node {$s_{n_1}$}; 
		\draw (7,-1.5) node {$s_{n_2}$};
		\dobase{0}{0}
	\end{diagram} \cdots\,.
\end{align}
\begin{figure}[t]
	\centering
	\includegraphics[width=\linewidth]{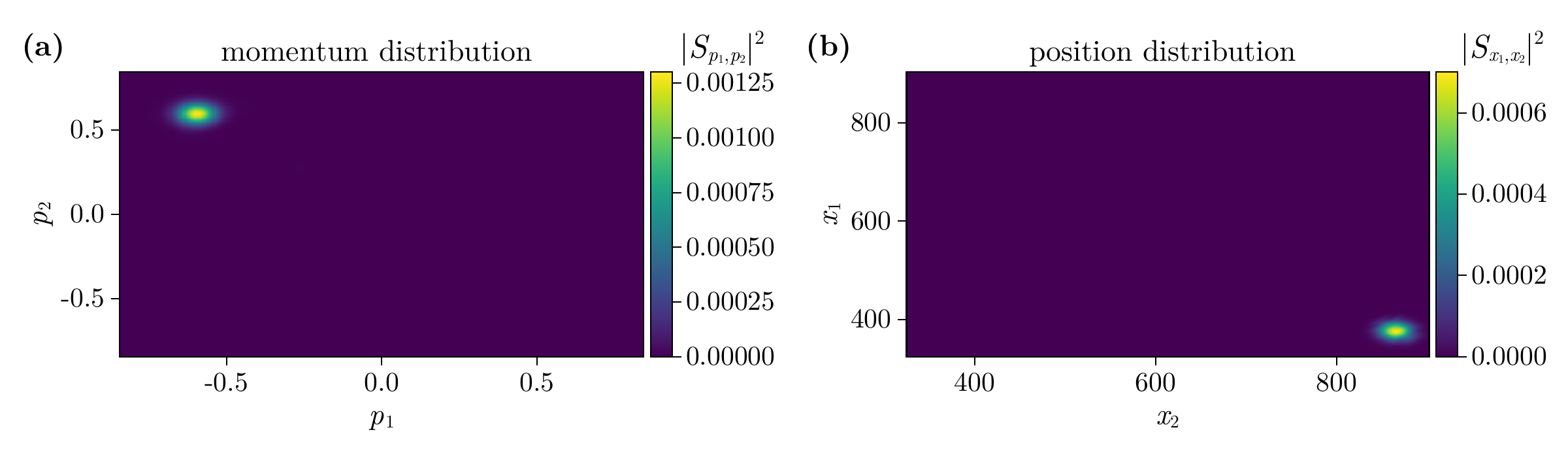}
	\caption{(a) Momentum and (b) position probability distributions of observing a $\pi_1\pi_1$ meson pair at $t=345$ following the $\pi_1\pi_1$ collision at $\varepsilon=0.07$ in Fig.~3 of the main text.}
	\label{fig:meson-S-matrix}
\end{figure}
The sums are restricted such that the two excitation tensors $B$ appear in disjoint regions $W_L$ and $W_R$ and are separated by some minimum number $r$ of vacuum tensors, ensuring that there are no interactions between the particles (we found $r=40$ to be sufficient for both meson and quark scattering in the main text).
Moreover, since we are projecting on a time-evolved state where the particles are constrained to a finite window of the uMPS, we further restrict the sums to the sites within this nonuniform window.
Using the left gauge-fixing condition for the left-most $B$ tensor, and the right gauge-fixing condition for the right-most $B$ tensor~\cite{vanderstraetenTangentspaceMethodsUniform2019}, ensures that overlaps where either of the $B$ tensors is outside of the window are exactly zero.
The finite sums in \cref{eq:two-particle-basis} can then be summed exactly and expressed as a single MPS, similarly to \cref{eq:wavepacket-summed}.

To compute the full $n$-particle momentum distribution (for $n=2$, $S_{p_1,p_2}\equiv \bra{\Phi_{p_1,p_2}}\ket{\psi(t)}$, where $\ket{\psi(t)}$ is the time-evolved state), $M_p^n$ contractions need to be performed, where $M_p$ is the desired number of $p$ samples (i.e., the equivalent of $N_p$ in the incoming wave-packet construction).
This can be sped up by precomputing partial contractions. For example, for the two-particle case, we compute the partial left and right contractions for each value of $p$, which scale as $\sim M_p |W_L|$ and $\sim M_p|W_R|$, respectively. 
Then, the $\sim M_p^2$ contractions are very efficient since the remaining tensor network is of length $\sim 1$. 
As an example, \cref{fig:meson-S-matrix} depicts the result of projecting the state following the meson-meson collision (this process is depicted in Fig.~3 of the main text) on the lightest meson-meson ($\pi_1\pi_1$) two-particle basis. 
From the momentum distribution, the position distribution ($S_{x_1,x_2}$) is computed by a Fourier transform. For this to work, it is crucial that the phases of the $B(p)$ tensors are fixed prior to the projection, as described in \cref{sec:wavepackets_prep}. To avoid multiple unphysical copies due to the periodicity of the discrete Fourier transform, we choose the number of $p$ samples, $M_p$, to be larger than the window size (we used $M_p=400$ for the quark-antiquark scattering and $M_p=1300$ for the meson-meson scattering).
Finally, fitting Gaussians to the marginals of these distributions produces the information plotted in Fig.~3(c) of the main text.

\FloatBarrier